\documentclass[epj,nopacs]{svjour}

\usepackage[utf8]{inputenc}
\usepackage[english]{babel}
\usepackage{bbm}
\usepackage{nicefrac}
\usepackage{slashed}
\usepackage{pstricks}
\usepackage{graphicx}
\usepackage{hyperref}
\usepackage{leftidx}
\usepackage{environ}
\usepackage{mathtools}
\usepackage{xspace}
\usepackage{suffix}
\usepackage{url}
\usepackage{slashed}
\usepackage{scalerel,stackengine}
\usepackage{array}
\usepackage{cuted}
\usepackage{isotope}

\newcommand{\ie}{\textit{i.e.}\xspace}

\newcommand{\apriori}{\textit{a priori}\xspace}
\WithSuffix\newcommand\apriori*{\textit{a-priori}\xspace}

\newcommand{\via}{\textit{via}\xspace}

\newcommand{\mathspace}{\ \ }
\newcommand{\mathtext}[1]{\mathspace\text{#1}\mathspace}

\newcommand{\fm}{\ensuremath{\mathrm{fm}}}

\newcommand{\vecr}{\mathbf{r}}
\newcommand{\vecx}{\mathbf{x}}

\newcommand{\vecp}{\mathbf{p}}
\newcommand{\veck}{\mathbf{k}}
\newcommand{\vecq}{\mathbf{q}}

\newcommand{\vecu}{\mathbf{u}}
\newcommand{\vecv}{\mathbf{v}}

\newcommand{\vecR}{\mathbf{R}}

\newcommand{\vdelta}{\delta^{(3)}}

\newcommand{\dd}{\mathrm{d}}

\newcommand{\ii}{\mathrm{i}}

\newcommand{\vcalD}{\boldsymbol{\cal D}}

\newcommand{\OO}{\mathcal{O}}

\newcommand{\YY}{Y}
\newcommand{\YYY}{\scalebox{1.33}{\ensuremath{\mathcal{Y}}}}

\newcommand{\one}{\mathbbm{1}}

\newcommand{\diag}{\mathrm{diag}}

\newcommand{\mean}[1]{\langle #1\rangle}

\newcommand{\bra}[1]{\langle #1|}

\newcommand{\ket}[1]{|#1\rangle}

\newcommand{\braket}[2]{\langle #1|#2\rangle}

\newcommand{\mbraket}[3]{\langle #1|#2|#3\rangle}

\newcommand{\mbraketred}[3]{\langle #1|\;\!\!|#2|\;\!\!|#3\rangle}

\newcommand{\abs}[1]{\left|#1\right|}

\newcolumntype{C}{>{$}p{1em}<{$}}

\newcommand{\SixJ}[6]{\left\{\!\!%
\begin{array}{CCC}%
{#1} & {#2} & {#3}\\[0.2em]%
{#4} & {#5} & {#6}%
\end{array}%
\!\!\right\}%
}

\newcommand{\couple}[3]{\left({#1}{#2}\right)\!{#3}}

\newcommand{\MN}{M_N}

\newcommand{\Mpi}{M_\pi}

\newcommand{\Mhi}{M_{\text{hi}}}

\newcommand{\ThreeSOne}{\ensuremath{{}^3S_1}\xspace}
\newcommand{\OneSNot}{\ensuremath{{}^1S_0}\xspace}

\newcommand{\TwoH}{\ensuremath{{}^2\mathrm{H}}\xspace}

\newcommand{\Triton}{\ensuremath{{}^3\mathrm{H}}\xspace}
\newcommand{\ThreeH}{\Triton}
\newcommand{\ThreeHe}{\ensuremath{{}^3\mathrm{He}}\xspace}
\newcommand{\FourHe}{\ensuremath{{}^4\mathrm{He}}\xspace}

\newcommand*\rvec[1]%
{\ensuremath{\overset{\smash{\raisebox{-1.5pt}{\tiny$\rightarrow$}}}{#1}}}
\newcommand*\lvec[1]%
{\ensuremath{\overset{\smash{\raisebox{-1.5pt}{\tiny$\leftarrow$}}}{#1}}}

\newcommand{\Bgen}{\mathcal{B}}

\newcommand{\MeV}{\ensuremath{\mathrm{MeV}}}

\newcommand{\sti}{\mathbf{i}}

\newcommand{\LO}{\text{LO}\xspace}
\newcommand{\NLO}{\text{NLO}\xspace}
\newcommand{\NNLO}{\text{N$^2$LO}\xspace}

\NewEnviron{subalign}[1][]{%
\begin{subequations}\begin{align}
  \BODY
\end{align}\label{#1}\end{subequations}
}

\NewEnviron{spliteq}{%
\begin{equation}\begin{split}
  \BODY
\end{split}\end{equation}
}

\stackMath
\newcommand\reallywidehat[1]{%
\savestack{\tmpbox}{\stretchto{%
  \scaleto{%
    \scalerel*[\widthof{\ensuremath{#1}}]{\kern-.6pt\bigwedge\kern-.6pt}%
    {\rule[-\textheight/2]{1ex}{\textheight}}
  }{\textheight}%
}{0.5ex}}%
\stackon[1pt]{#1}{\tmpbox}%
}

\def\makeheadbox{}

\begin{document}

\title{Energies and radii of light nuclei around unitarity}
\titlerunning{Energies and radii of light nuclei around unitarity}

\author{Sebastian König\inst{1}}
\authorrunning{Sebastian König}

\institute{%
Institut für Kernphysik,
Technische Universität Darmstadt,
64289 Darmstadt, Germany
\and
ExtreMe Matter Institute EMMI,
GSI Helmholtzzentrum für Schwerionenforschung GmbH,
64291 Darmstadt, Germany\\
\email{sekoenig@theorie.ikp.physik.tu-darmstadt.de}
}

\abstract{
Light nuclei fall within a regime of universal physics governed by the
fact that the two-nucleon scattering lengths are large compared to the typical
nuclear interaction range set by one-pion exchange.
This places nuclear physics near the so-called unitarity limit in which
the scattering lengths are exactly infinite.
Effective field theory provides a powerful theoretical framework to capture
this separation of scales in a systematic way.
It is shown here that the nuclear force can be constructed as a perturbative
expansion around the unitarity limit and that this expansion has good
convergence properties for both the binding energies of $A=3,4$ nuclei as well
as for the radii of these states.
}

\date{\today}

\maketitle

\section{Effective field theory for systems near unitarity}
\label{sec:EFT}

Nuclear physics at very low energies hosts a fascinating emergent phenomenon:
out of the tremendously complicated dynamics of quarks and gluons,
governed by the strong interaction (Quantum Chromodynamics, QCD) that is highly
nonperturbative in this regime, ultimately arise strikingly simple features for
systems of few nucleons.
It was realized many decades
ago~\cite{Schwinger:1947xx,Barker:1949zz,Chew:1949zz,Bethe:1949yr} that the
low-energy two-nucleon system can be parametrized with a formula that has
become famous, in nuclear physics and beyond, as the \emph{effective range 
expansion (ERE)}:
\begin{equation}
 k\cot\delta_0(k) = {-}\frac{1}{a} + \frac{r}{2} k^2 + \cdots \,.
\label{eq:ERE}
\end{equation}
Here $\delta_0(k)$ denotes the $S$-wave scattering phase shift for two
particles (here, nucleons in a single fixed spin configuration) with relative
momentum $k$.
The leading parameter in this expansion, called ``scattering length'' and
denoted by $a$, governs the nucleon-nucleon ($N\!N$) cross section at low
energies and completely determines it in the limit where the relative momentum
$k$ of the nucleons goes to zero:
\begin{equation}
 \sigma = 4\pi a^2 + \OO(k^2) \,.
\label{eq:sigma-a}
\end{equation}
Empirically, in the \ThreeSOne (``$t$'') and \OneSNot (``$s$'') $N\!N$ spin
channels the values are known to be $a_t\simeq 5.4~\fm$ and $a_s\simeq
{-}23.7~\fm$, respectively.
Compared to the typical range of the nuclear interaction, set by one-pion
exchange providing the longest-range component as $R\sim \Mpi^{-1} \simeq
1.4~\fm$, these scattering lengths are unnaturally large, $a_{s,t}\gg R$.
Through Eq.~\eqref{eq:sigma-a} this implies that the nuclear force is
particularly strong as the energy goes to zero.
The fact that this is so can be understood as an accidental ``fine tuning'' of
the QCD parameters~\cite{Beane:2001bc,Beane:2002vs,Epelbaum:2002gb,%
Beane:2002xf,Braaten:2003eu} (the quark masses, in particular) to be close to a
critical point where the scattering lengths diverge.
This point is called the ``unitarity (or unitary) limit,'' and it is the heart
of the emergent simplicity mentioned at the outset.

Systems near the unitarity limit exhibit universal features.
As the two-body scattering length becomes large, the details of the underlying
interaction largely cease to matter and to a very good approximation the
behavior of the system is determined qualitatively by the fact that $a$ is
large, and quantitatively by how large exactly it is.
This phenomenon places low-energy nuclear systems in a common universality
class with other systems near unitarity, such as cold atomic gases (where the
scattering length can be tuned experimentally \via Feshbach
resonances~\cite{Chin:2010xx}), or certain mesons which can be
interpreted as hadronic molecules~\cite{Braaten:2003he}.

In the two-body sector, universality relates scattering parameters to shallow
bound and virtual states.
This is a consequence of Eq.~\eqref{eq:ERE} and the principle of analyticity:
the ERE provides an expansion of the $S$ matrix, so whenever poles at complex
momenta---in particular bound and virtual states, which reside at purely 
imaginary momenta---fall within the radius of convergence of the expansion, they
are described by the same parameters.
Since, schematically, the $S$-wave $S$ matrix is given by $1 + \ii t$ with
\begin{equation}
 t(k) \sim \frac{1}{k\cot\delta_0(k) -\ii k} \,,
\end{equation}
the pole condition is $\cot\delta_0(k) = \ii$ for some $k = \ii\kappa$.
Keeping only the first term in the ERE this gives $\kappa = 1/a$ and one sees
that for positive (negative) $a$ one has a bound (virtual) state in the system.
Having a sufficiently large $\abs{a}$ ensures that indeed these momentum scales
lie within the radius of convergence of the ERE, which for nucleons is
determined by the position of the pion cut, $\Mpi/2$.
For the two-nucleon system at the physical point one has the deuteron as a
shallow bound state ($B_D\simeq 2.224~\MeV$, with the difference to $1/(\MN
a_t^2) \approx 1.41~\MeV$ being due to range corrections) in the \ThreeSOne
channel, and a very shallow virtual state at $B_{N\!N^*}\simeq 0.068~\MeV$ (with
a relatively small range correction since $\abs{a_s}$ is so large).
In the unitarity limit, $a_{s,t}\to\infty$, both of these poles
become zero-energy $S$-wave states.

A more striking universal behavior is encountered for three and more particles:
in the unitarity limit there exists an infinite tower of three-body states,
geometrically spaced (the binding energy of each subsequent state is
given by a fixed factor times the previous level) and accumulating at zero 
energy, a phenomenon that has become famous as Efimov 
effect~\cite{Efimov:1970zz}.
At large but finite scattering length the spectrum is cut off in the infrared
due to the existence of a two-body pole in the $S$ matrix.
It was shown in Refs.~\cite{Bedaque:1998kg,Bedaque:1998km,Bedaque:1999ve} that
for physical values of the $N\!N$ scattering lengths the triton can be
interpreted as the single remaining bound state of such an Efimov tower.
More recently it was established in a model-independent way~\cite{Rupak:2018gnc}
that a virtual state in the three-nucleon ($3N$) system, known to exist for a
long time~\cite{vanOers:1967lny,Girard:1979zza}, is in fact an $S$-matrix pole
that would be an excited Efimov state instead if nature were just slightly
closer to the unitarity limit.
This confirms a relation previously observed in a separable potential
model~\cite{Adhikari:1982zzb}.

The phenomenon continues at the four-body level.
At unitarity, each three-boson Efimov state (with binding energy $B_3$) is
associated with two four-boson states~\cite{Hammer:2006ct}.
One of these is almost five times as deeply bound as the trimer, $B_4/B_3\simeq
4.611$, while the other resides just below the particle-trimer threshold,
$B_{4^*}/B_3\simeq 1.002$~\cite{Deltuva:2010xd}.
Universality implies that if the $N\!N$ scattering lengths were infinite, the
ground state of \isotope[4]{He} would be located at $4.6$ times the binding
energy of the triton (neglecting Coulomb and other isospin breaking effects).
In nature, the ground state is at $B_\alpha/B_H\simeq 3.66$, and there exists
a $0^+$ resonance state just above the proton-triton threshold, \ie, one has
$B_{\alpha^*}/B_H\simeq 1.05$, where $B_H\simeq 7.72~\MeV$ is the \ThreeHe
binding energy, taken as reference here to at least partially account for 
isospin breaking effects.
The closeness of these ratios to the unitarity-limit values is a strong
indication that nature may be perturbatively close to unitarity for systems of
at least four nucleons.

In the following, this work discusses how to construct an effective field theory
(EFT) that captures all phenomena mentioned above.
EFTs are a powerful tool widely used in modern theoretical physics.
In nuclear physics they enable the consistent construction of nuclear forces
systematically connected to QCD by choosing a ``theoretical resolution'' at 
which effective interactions between degrees of freedom appropriate for the 
energy scales of interest are constructed.
The richness of nuclear phenomena implies that there are a number of different
EFTs relevant for nuclear physics, forming the ``tower'' of theories that
gives rise to the name of the topical issue this work is contributed to.
A recent review of EFTs that use nucleons and mesons as degrees of freedom
can be found in Ref.~\cite{Hammer:2019poc}.
Here the focus is on setting up an EFT that systematically expands light nuclei
around the unitarity limit, expanding on previous work considering the unitarity
expansion~\cite{Konig:2016utl,Konig:2016iny} by considering charge radii of
light nuclei in addition to binding energies.
After the setup and implementation of the expansion as well as results are
presented in the following sections, Sec.~\ref{sec:Outlook} will come back to
the question where the unitarity expansion resides within the tower, or 
landscape, of nuclear EFTs.

\medskip
As a variant of what has become known as ``Pionless EFT,'' the unitarity
expansion is defined in terms of a Lagrange density
\begin{spliteq}
 \mathcal{L} &= N^\dagger\left(\ii {\cal D}_0+\frac{\vcalD^2}{2\MN}\right)N \\
 &+ \sum\nolimits_{\sti}C_{0,\sti}
 \left(N^T P_{\sti} N\right)^\dagger \left(N^T P_{\sti} N\right)
 + D_0 \left(N^\dagger N\right)^3
 + \cdots \,,
\label{eq:L}
\end{spliteq}
where the notation has been taken over from
Refs.~\cite{Konig:2015aka,Konig:2016utl}.
The degrees of freedom are nonrelativistic nucleon isospin doublets
$N=(p\;n)^T$, coupled to photon fields $A_\mu$ \via the covariant
derivative $\mathcal{D}_\mu = \partial_\mu + \ii eA_\mu (1+\tau_3)/2$, where
$e$ is the proton charge and $\tau_a$ is used to label isospin Pauli matrices.
Of the electromagnetic interactions, only the static Coulomb potential is
relevant up to high orders (defined later), where corrections from transverse
photons will eventually enter.
The strong interaction is parametrized in Eq.~\eqref{eq:L} by the ``low-energy
constants (LECs)'' $C_{0,\sti}$ and $D_{0}$, defining contact
(zero-range) interactions without derivatives between two and three nucleons,
respectively.
The $P_{\sti}$ denote projectors onto the $N\!N$ $S$ waves, $\sti =
\OneSNot,\ThreeSOne$.
Contact interactions with derivatives as well as higher-body forces are
contained in the ellipses in Eq.~\eqref{eq:L}, along with other interactions
not shown explicitly here.

The ellipses in Eq.~\eqref{eq:L} represent a fundamental feature of
an EFT, namely that the Lagrangian contains all possible terms which are 
allowed by the symmetries of the system at hand.
For the EFT of nucleons considered here these symmetries are inherited
from QCD as the underlying theory: each term in Eq.~\eqref{eq:L} is required
to be invariant under Galilean boosts (plus systematic relativistic
corrections), rotations, isospin, and other discrete symmetries.
It is of course not arbitrary that Eq.~\eqref{eq:L} explicitly shows
some terms but not others.
In order to be predictive, each EFT comes with an organizational principle
called ``power counting,'' which attributes the various terms to increasingly
higher orders.
A starting point for this organization is typically a naïve dimensional analysis
(NDA): fields and derivatives acting on them are assigned their canonical
dimensions, defining the exponent of a typical low-momentum scale $Q$.
In order for each term to have overall dimension four, appropriate powers of
the EFT breakdown scale $\Mhi$ are included in the prefactor.
For the standard Pionless EFT expansion, $\Mhi \sim R^{{-}1}\sim\Mpi$, and
this is kept for the construction of the unitarity expansion.
However, while standard Pionless EFT assumes $Q \sim 1/a_{s,t}$,
Ref.~\cite{Konig:2016utl} suggested to count these scales separately
as $\aleph \sim 1/a_{s,t}$ while assuming that
\begin{equation}
 Q \sim Q_A = \sqrt{2M_NB_A/A} \,.
\label{eq:Q-A}
\end{equation}
This is a momentum scale associated with the binding energy per nucleon in an
$A$-nucleon system, which for $A=2$ coincides with the canonical definition of
the binding momentum.
With this assumption one obtains $\aleph < Q < 1/R$ such that it is possible to
set up a \emph{combined} expansion in two parameters $\aleph/Q$ and $QR$.
Coulomb effects are perturbative for momenta of order $Q_A$ as well and
naturally captured by the expansion if one takes into account that the Coulomb
momentum scale $k_C = \alpha\MN$ with the fine-structure constant
$\alpha\approx1/137$, is naturally included in the $\aleph$
scale~\cite{Konig:2015aka}.

For a calculation of few-body states it is convenient to switch from the
Lagrangian formulation to standard quantum mechanics expressed in terms of
potentials.
In the two-body sector, it possible to write
\begin{equation}
 V_{2,\sti}^{(0)} = C_{0,\sti}^{(0)}\ket{g}\bra{g} \,,
\label{eq:V2-C0}
\end{equation}
where $C_{0,\sti}^{(0)}$ is the leading-order (LO) piece of the non-derivative
contact LEC $C_{0,\sti}$ in Eq.~\eqref{eq:L}, which has an expansion of the form
\begin{equation}
 C_{0,\sti} = C_{0,\sti}^{(0)} + C_{0,\sti}^{(1)} + \cdots \,.
\end{equation}
Apart from this, $V_{2,\sti}^{(0)}$ is defined in terms of a separable Gaussian
regulator function, given by
\begin{equation}
 \braket{p}{g} = g(p^2) = \exp({-}p^2/\Lambda^2)
\label{eq:g-2}
\end{equation}
in momentum space.
This makes the zero-range theory well defined by regularizing the otherwise 
divergent interaction \via the introduction of a cutoff scale $\Lambda$.
Both the value of $\Lambda$ and the particular form of the regulator function
are arbitrary and renormalization, discussed below, ensures that observables
are independent of these choices.
The separable form is however particularly convenient for the formalism
explained in the following.
It makes it possible to algebraically solve the
Lippmann-Schwinger equations for the LO $T$ matrices $t_\sti^{(0)}$,
\begin{equation}
 t_\sti^{(0)} = V_{2,\sti}^{(0)} + V_{2,\sti}^{(0)} G_0 t_\sti^{(0)} \,,
\label{eq:LS-t0}
\end{equation}
where
\begin{equation}
 G_0(z) = \frac{1}{z - H_0}
\end{equation}
defines, for an arbitrary energy $z$, the two-body Green's function in terms of
the free (purely kinetic) Hamiltonian $H_0$.
The result is:
\begin{align}
 t_\sti^{(0)}(z;\veck,\veck')
 &= \mbraket{\veck}{t_\sti^{(0)}}{\veck'} = g(k^2) \tau_\sti(z) g(k'^2) \,, \\
 \tau_\sti(z) &= \left[1/{C_{0,\sti}^{(0)}} - \mbraket{g}{G_0}{g}\right]^{-1}
 \,.
\label{eq:T-LO}
\end{align}
The regulator ensures that $\mbraket{g}{G_0}{g}$ is finite.
At the on-shell point, $E = k^2/\MN$ and $\veck=\veck'$, this solution can be
matched directly to the ERE, yielding
\begin{equation}
 C_{0,\sti}^{(0)} \to C_{0,\sti}^{(0)}(\Lambda)
 = \frac{1}{2\pi^2\MN}\left(\frac{1}{a_\sti} -\theta_0\Lambda\right)^{{-}1} \,.
\label{eq:C0-std}
\end{equation}
With this running coupling appearing in Eq.~\eqref{eq:V2-C0}, the two-body
sector of the theory is renormalized.
The number $\theta_0$ in general depends on the choice of regulator; for the
Gaussian form used here one finds $\theta_0 = 1/\sqrt{2\pi}$.
In the unitarity limit, $1/a_\sti = 0$, such that the leading-order two-body
interaction is parameter free:
\begin{equation}
 C_{0,\sti}^{(0)}(\Lambda)
 = \frac{{-}1}{2\pi^2\MN}\frac{1}{\theta_0\Lambda} \,.
\label{eq:C0-U}
\end{equation}

Perturbative higher orders are defined by formally expanding the full $T$
matrices $t_\sti$ as
\begin{equation}
 t_\sti = t_\sti^{(0)} + t_\sti^{(1)} + t_\sti^{(2)} + \cdots \,,
\end{equation}
where $t_\sti^{(0)}$ is defined by Eq.~\eqref{eq:T-LO}.
The corrections $t_\sti^{(n)}$ for $n>0$ can conveniently be obtained by
solving similar integral equations~\cite{Vanasse:2013sda,Konig:2016iny}.
For the unitarity expansion, corrections from the finite scattering length
enter at NLO \via $C_{0,\sti}^{(1)}$, yielding a separable potential
$V_{2,\sti}^{(1)}$ with the same form as Eq.~\eqref{eq:V2-C0}.
For $t_\sti^{(1)}$, this gives rise to
\begin{equation}
 t_\sti^{(1)} = V_{2,\sti}^{(1)}
 + V_{2,\sti}^{(1)} G_0 t_\sti^{(0)}
 + V_{2,\sti}^{(0)} G_0 t_\sti^{(1)} \,,
\label{eq:LS-t1}
\end{equation}
which, just like the LO equation, can be solved algebraically (see
Ref.~\cite{Konig:2016iny} for explicit details).
From this procedure one obtains
\begin{equation}
 C_{0,\sti}^{(1)}(\Lambda)
 = \frac{{-}2\pi^2\MN}{a_\sti} \left[C_{0,\sti}^{(0)}(\Lambda)\right]^2 \,.
\label{eq:C0-U-1}
\end{equation}

Range corrections enter at NLO together with $C_{0,\sti}^{(1)}$.
They are generated by contact interactions involving qua\-dratic
de\-ri\-va\-tives acting on the nucleon fields, included in the ellipses in
Eq.~\eqref{eq:L}.
The corresponding potential can be written in momentum space as
$C^{(1)}_{2,\sti} g(k^2) \left(k^2+k'^2\right) g(k'^2)$.
By virtue of this still being a separable interaction, the corresponding version
of Eq.~\eqref{eq:LS-t1} with
\begin{equation}
 \mbraket{\vec{k}}{V^{(1)}_{2,\sti}}{\vec{k}'}
 = C_{0,\sti}^{(1)} g(k^2) g(k'^2)
 + C^{(1)}_{2,\sti} g(k^2) \left(k^2+k'^2\right) g(k'^2)
\end{equation}
can still be solved algebraically.
Matching the result to the ERE~\eqref{eq:ERE} up to the quadratic term gives
\begin{subalign}
 C_{2,\sti}^{(1)}(\Lambda)
 &= {\pi^2\MN} \left(\frac{r}{2} - \frac{1}{\theta_2\Lambda}\right)
 \left[C_{0,\sti}^{(0)}(\Lambda)\right]^2 \,,
 \\
 C_{0,\sti}^{(1)}(\Lambda)
 &= {4\pi^2\MN} {\theta_2\Lambda^3}
 C_{0,\sti}^{(0)}(\Lambda) C_{2,\sti}^{(1)}(\Lambda)
 \,,
\end{subalign}
with $\theta_2 = \theta_0/4$ for the Gaussian regulator used here.
Going to higher orders is straightforward, proceeding in the same way \via
integral equations that can be solved
algebraically, recursively using the
solutions of previous orders~\cite{Vanasse:2013sda,Konig:2016iny}.
At second order, the $T$-matrix correction is obtained from
\begin{equation}
 t_\sti^{(2)} = V_{2,\sti}^{(2)}
 + V_{2,\sti}^{(1)} G_0 t_\sti^{(1)}
 + V_{2,\sti}^{(0)} G_0 t_\sti^{(2)} \,.
\label{eq:T-N2LO}
\end{equation}
For a perturbative treatment of Coulomb contributions, which are neglected
in this work, see Refs.~\cite{Konig:2015aka,Konig:2016iny}.

Leading order is however not complete with only the $C_{0,\sti}^{(0)}$
interactions.
It is a distinct feature of Pionless EFT, intimately related to the Efimov
effect~\cite{Bedaque:1998kg,Bedaque:1998km,Bedaque:1999ve}, that a three-nucleon
interaction enters at LO.
Naïvely it would be expected to contribute only much later in the power counting
because the larger number of fields, according to NDA, implies more inverse
powers of $\Mhi$ in the prefactor.
Analogously to the two-body interactions, the potential induced by the term
involving $D_0$ in Eq.~\eqref{eq:L} can be written in a separable form,
\begin{equation}
 V_3^{(0)} = D_0^{(0)} \, \ket{\Triton}\ket{\xi}\bra{\xi}\bra{\Triton}
\label{eq:V3-0}
\end{equation}
at LO, where $\ket{\Triton}$ projects onto a $J=T=\nicefrac12$ three-nucleon 
state and the regulator $\ket{\xi}$ is defined, for Jacobi momenta $\vecu_1 =
\frac12(\veck_1-\veck_2)$ and $\vecu_2 =
\frac23[\veck_3-\frac12(\veck_1+\veck_2)]$, as
\begin{equation}
 \braket{\vecu_1 \vecu_2}{\xi} = g\big(u_1^2+\frac{3}{4}u_2^2\big) \,.
\end{equation}
The $\veck_i$ label the individual nucleon momenta.
An NLO correction $V_3^{(1)}$ has the same form as Eq.~\eqref{eq:V3-0}, but
involves the LEC $D_0^{(1)}$.
Both $D_0^{(0)}$ and $D_0^{(1)}$ are determined by the triton binding energy
and then enter in other calculations of $A\geq3$ observables.
These are described in the following.

\section{Faddeev- and Faddeev-Yakubovsky equations}
\label{sec:Faddeev}

This section gives an overview of the three- and four-body formalism, 
implementing a unified framework to solve, respectively, the Faddeev and 
Faddeev-Yakubovsky used to obtain the results presented in the following 
sections.
The main aspects are explained in broad strokes, referring the reader to the 
references given for more background.
Developments needed to calculate charge radii along with perturbative 
corrections, are, however elaborated on a further, with key results explained 
the main text and additional details provided in 
Appendices~\ref{sec:ChargeOperators} and~\ref{sec:PerturbationTheory}.

The basis for a description of the three-nucleon system are Jacobi momenta
\begin{subalign}[eqs:Jacobi-3]
 \vecu_1 &=
 \frac12(\veck_1-\veck_2) \,, \\
 \vecu_2 &=
 \frac23[\veck_3-\frac12(\veck_1+\veck_2)] \,,
\end{subalign}
where the $\veck_i$ label the individual nucleon momenta, conjugate to position
vectors $\vecx_i$.
Projecting these momenta onto partial waves yields states $\ket{u_1u_2;s}$,
where
\begin{equation}
 \ket{s} = \ket{%
  \couple{l_2}{\couple{\couple{l_1}{s_1}{j_1}}{\tfrac12}{s_2}}{J};
  \couple{t_1}{\tfrac12}{T}
 }
\label{eq:s}
\end{equation}
collects angular momentum, spin, and isospin quantum numbers.
They are coupled such that $\couple{l_1}{s_1}{j_1}$ and $t_1$ describe the
two-nucleon subsystem, whereas $l_2$ denotes the orbital angular momentum
associated with the Jacobi momentum $u_2$ and $s_2$ is an intermediate quantum
number.
For the trinucleon bound states, the total spin and isospin are $J=T=1/2$.
These states are determined by solving the the Faddeev
equation~\cite{Stadler:1991zz}
\begin{multline}
 \ket{\psi^{(0)}} = G_0\,t^{(0)}\,P \ket{\psi^{(0)}} \\
  \null + \frac13(G_0 + G_0t^{(0)}\,G_0)V_3^{(0)}(1+P)\ket{\psi^{(0)}} \,,
 \label{eq:Faddeev-V3}
\end{multline}
where $\ket{\psi^{(0)}} = \ket{\psi_{(12)3}^{(0)}}$ is one of three equivalent
two-body Faddeev components.
As already done in the discussion of the two-body sector, explicit superscripts
``$(0)$'' are used to denote leading-order quantities.
Alternatively, one can incorporate the three-body interaction
$V_3^{(0)}$ by writing~\cite{Platter:2005}
\begin{subalign}[eq:Faddeev-t3]
 \ket{\psi^{(0)}} &= G_0\,t^{(0)}\,P \ket{\psi^{(0)}}
  + G_0\,t^{(0)}\,\ket{\psi_3^{(0)}} \,,
 \label{eq:Faddeev-t3-a} \\
 \ket{\psi_3^{(0)}} &= G_0\,t_3^{(0)}\,(1+P) \ket{\psi^{(0)}} \,,
 \label{eq:Faddeev-t3-b}
\end{subalign}
where $\ket{\psi_3^{(0)}}$ is an auxiliary amplitude, and
\begin{equation}
 t_3^{(0)} = V_3^{(0)} + V_3^{(0)}\,G_0\,t_3^{(0)} \,.
\end{equation}
In either form of the Faddeev equations, $G_0$ denotes the free three-body
Green's function and $P = P_{12}P_{23} + P_{13}P_{23}$ generates the
non-explicit components through permutations.
$t^{(0)}$ collectively denotes the two-body T-matrices $t_\sti^{(0)}$.
Note that $\ket{\psi_3^{(0)}}$ can be eliminated by inserting
Eq.~\eqref{eq:Faddeev-t3-b} into Eq.~\eqref{eq:Faddeev-t3-a}, yielding an
equation of the form
\begin{equation}
 \ket{\psi^{(0)}} = K^{(0)}\ket{\psi^{(0)}}
\end{equation}
with
\begin{equation}
 K^{(0)} = G_0\,t^{(0)}\,P + G_0\,t^{(0)}\,G_0\,t_3^{(0)}\,(1+P) \,,
\end{equation}
and alternatively a similar kernel can be obtained from
Eq.~\eqref{eq:Faddeev-V3} in terms of $V_3^{(0)}$.
Either form of the Faddeev equations is solved by representing it within the
space of states $\ket{u_1u_2;s}$, discretizing the momenta $u_{1,2}$ on a
quadrature mesh.
The binding energy is determined by varying the energy $E$, entering as an
argument to both $G_0$ and $t^{(0)}$, until the kernel $K$ has a unit
eigenvalue.
At that energy, $E = {-}B_0$, $\ket{\psi^{(0)}}$ can then be determined as the
corresponding eigenstate and finally one constructs the full wavefunction as
\begin{equation}
 \ket{\Psi^{(0)}} = (1 + P) \ket{\psi^{(0)}} + \ket{\psi_3^{(0)}} \,.
\label{eq:Psi-3}
\end{equation}
If one starts from Eq.~\eqref{eq:Faddeev-V3}, there is no explicit three-body
Faddeev component and one simply has $\ket{\Psi^{(0)}} = (1 + P)
\ket{\psi^{(0)}}$.

Fundamentally, the permutation operator $P$ leads to a coupling of
different partial waves (see Ref.~\cite{Gloeckle:1983} for an excellent 
pedagogical discussion of both this and the Faddeev equations in general), and 
for the construction of the full wavefunction~\eqref{eq:Psi-3} it is important 
to include higher partial waves: the proper antisymmetry of
$\ket{\Psi^{(0)}}$ is only recovered as more and more states are included,
which means that in principle all observables calculated from $\ket{\Psi^{(0)}}$
have to be checked for convergence.

The full wavefunctions $\ket{\Psi^{(0)}}$ are used to calculate both
perturbative shifts for the binding energy,
\begin{equation}
 B_1 = \mbraket{\Psi^{(0)}}{V^{(1)}}{\Psi^{(0)}} \,,
\label{eq:B1}
\end{equation}
as well as the radius at leading order (see Sec.~\ref{sec:Radii}).
Note that $\ket{\Psi^{(0)}}$ is assumed here to be properly normalized,
\begin{equation}
 \braket{\Psi^{(0)}}{\Psi^{(0)}} = 1 \,.
\label{eq:N}
\end{equation}
In calculating the matrix elements in Eqs.~\eqref{eq:B1} and~\eqref{eq:N} it is
advantageous to exploit the antisymmetry of $\ket{\Psi^{(0)}}$ as much as 
possible because that will speed up convergence of results with respect to the
number of partial-wave channels.
For example, the two-body part of the full potential $V^{(1)}$ can be
expressed through permutations in terms of only the potential between the pair
of nucleons 1 and 2, and it holds that $(1+P)(1+P) = 3(1+P)$.

Note furthermore that Eqs.~\eqref{eq:Faddeev-t3}, and similarly
Eq.~\eqref{eq:Faddeev-V3}, can be significantly simplified by exploiting the
fact that the two- and three-body interactions are separable and act only within
$S$ waves.
As a result, it suffices to work with merely two coupled equations for the
triton~\cite{Platter:2005}, and using the procedure described in the previous
paragraph it is furthermore possible to eliminate all intermediate higher 
partial-wave components int the calculation of $B_1$.
These simplifications are used in the practical implementation to fit
the three-body LECs $D_0^{(0)}$ and $D_0^{(1)}$.

\medskip
Describing the four-nucleon system requires an additional Jacobi momentum
\begin{equation}
 \vecu_3 =
 \frac34[\veck_4-\frac13(\veck_1+\veck_2+\veck_3)] \,,
\label{eq:Jacobi-4-u}
\end{equation}
as well as an alternative set of momenta $(\vecv_1,\vecv_2,\vecv_3)$,
describing a 2+2 cluster setup, \ie, $\vecv_1=\vecu_1$, $\vecv_3$ denotes the
relative momentum in the (34) system, and $\vecv_2$ is defined as the relative
momentum between the (12) and (34) subsystems.

\begin{strip}
Including the remaining quantum numbers, this leads to channel states
\begin{subalign}[eqs:Coupling-4]
 \ket{a} &= \ket{\!
   \couple{l_2}{\couple{\couple{l_1}{s_1}{j_1}}{\tfrac12}{s_2}}{j_2},
   \couple{l_3}{\tfrac12}{j_3},
   \couple{j_2}{j_3}{J};
   \couple{\couple{t_1}{\tfrac12}{t_2}}{\tfrac12}{T}
 } \,,
 \label{eq:a}
 \\
 \ket{b} &= \ket{\!
  \couple{\lambda_2}{\couple{\lambda_1}{\sigma_1}{\iota_1}}{\iota_2},
  \couple{\lambda_3}{\sigma_3}{\iota_3},
  \couple{\iota_2}{\iota_3}{J};
  \couple{\tau_2}{\tau_3}{T}
 } \,.
 \label{eq:b}
\end{subalign}
\end{strip}
The $\ket{a}$ are a straightforward extension of three-nuc\-le\-on
states~\eqref{eq:s}, including the angular momentum $l_3$ associated with
$\vecu_3$ as well as spin and isospin $\tfrac12$ for the fourth nucleon.
For the $b$ states, $(\lambda_1,\sigma_1,\tau_1)$ and
$(\lambda_3,\sigma_3,\tau_3)$ are two-body quantum numbers for the $(12)$ and
$(34)$ subsystems, respectively, whereas $\lambda_{1,2,3}$ are the angular
momenta associated with $v_{1,2,3}$.
For \FourHe the total spin and isospin are $J=T=0$.

Following Refs.~\cite{Platter:2005,Kamada:1992aa}, the Faddeev-Yakubovsky
equations can be written as
\begin{subalign}[eq:FaddeevYakubovsky]
 \ket{\psi_A^{(0)}} &= G_0 t^{(0)} P
 \big[(1 - P_{34})\ket{\psi_A^{(0)}} + \ket{\psi_B^{(0)}}\big] \\
 \nonumber &\hspace{6em}
 + \frac13(1 + G_0 t^{(0)}) G_0 V_3^{(0)} \ket{\Psi^{(0)}}
 \\
 \ket{\psi_B} &= G_0 t^{(0)} \tilde{P}
 \big[(1 - P_{34})\ket{\psi_A^{(0)}} + \ket{\psi_B^{(0)}}\big] \,,
\end{subalign}
corresponding to the decomposition
\begin{multline}
 \ket{\Psi^{(0)}} = (1 + P)\Big[(1 - P_{34} - P_{34}P)\ket{\psi_A^{(0)}} \\
  \null + (1 + \tilde{P})\ket{\psi_B^{(0)}}\Big]
\label{eq:Psi-0-4B}
\end{multline}
of the full four-body wavefunction.
The two distinct Fadde\-ev-Yakubovsky components $\ket{\psi_A}$ and
$\ket{\psi_B}$ correspond to, respectively, 3+1 and 2+2 cluster configurations
of the four-body system, with the former naturally expressed in terms of the
Jacobi momenta $u_i$ and states $\ket{a}$, and the latter in terms of $v_i$ and
$\ket{b}$.
Note that the same notation is used here as for the three-body case, and
$G_0$ in Eqs.~\eqref{eq:FaddeevYakubovsky} now represents the free
four-body Green's function.
In addition to the operator $P$ already encountered in the three-body
system, Eqs.~\eqref{eq:FaddeevYakubovsky} include the further permutations
$P_{34}$ and $\tilde{P} = P_{13}P_{24}$ to ensure proper antisymmetry.
In fact, the overall symmetry is determined by the sign in front of $P_{34}$: to
study a bosonic system, one would use $(1+P_{34})$ acting on
$\ket{\psi_A}$ to construct fully symmetric states.

The structure of Eqs.~\eqref{eq:FaddeevYakubovsky} can be made clearer by
rewriting them in a generic matrix form,
\begin{equation}
 \left(\one - \hat{K}^{(0)}\right) \ket{\boldsymbol{\psi}^{(0)}} = 0 \,,
\label{eq:FY-0-mat}
\end{equation}
with $\ket{\boldsymbol{\psi}^{(0)}}
= (\ket{\psi_A^{(0)}}, \ket{\psi_B^{(0)}})^T$ and the kernel
\begin{equation}
 \hat{K}^{(0)} = G_0 t^{(0)} \hat{P}
  + \frac13(G_0 + G_0 t^{(0)} G_0) V_3^{(0)} \hat{P}_3 \,.
\label{eq:K-FY-0-23}
\end{equation}
In Eq.~\eqref{eq:K-FY-0-23}, $G_0$ and $t^{(0)}$ are understood to be
diagonal matrices, and the permutation operators are collected in
\begin{subalign}[eqs:P-hat]
 \label{eq:P-hat}
 \hat{P}_{\phantom{3}} &= \diag(P, \tilde{P})
 \otimes \begin{pmatrix}
  (1-P_{34}) & 1 \\
  (1-P_{34}) & 1
 \end{pmatrix} \,, \\
 \label{eq:P-hat-3}
 \hat{P}_{3} &= \begin{pmatrix}
  (1+P)(1-P_{34}-P_{34}P) & (1+P)(1+\tilde{P}) \\
  0 & 0
 \end{pmatrix} \,.
\end{subalign}
From this form, the structural analogy to the three-body Faddeev case becomes
obvious.

Just like for the Faddeev equations, Eqs.~\eqref{eq:FaddeevYakubovsky} are
solved by projecting onto states $\ket{u_1u_2u_3;a}$,
$\ket{v_1v_2v_3;b}$~\cite{Kamada:1992aa}, discretizing all momenta on a grid,
and looking for a unit eigenvalue of the resulting kernel matrix as a function
of the energy.
However, the set of coupled equations does not naturally truncate even
if all interactions are pure $S$ wave.
This means that already for a determination of the binding energies it is
necessary to truncate the sums in Eqs.~\eqref{eqs:Coupling-4} (by choosing all
total angular momenta $j_i$ and $\iota_i$ less than some $j_{\text{max}}$) and
study the numerical convergence of results as $j_{\text{max}}$ is increased.

\section{Binding energies of light nuclei}
\label{sec:Energies}

\begin{figure}[htbp]
\centering
\includegraphics[width=0.99\columnwidth,clip]{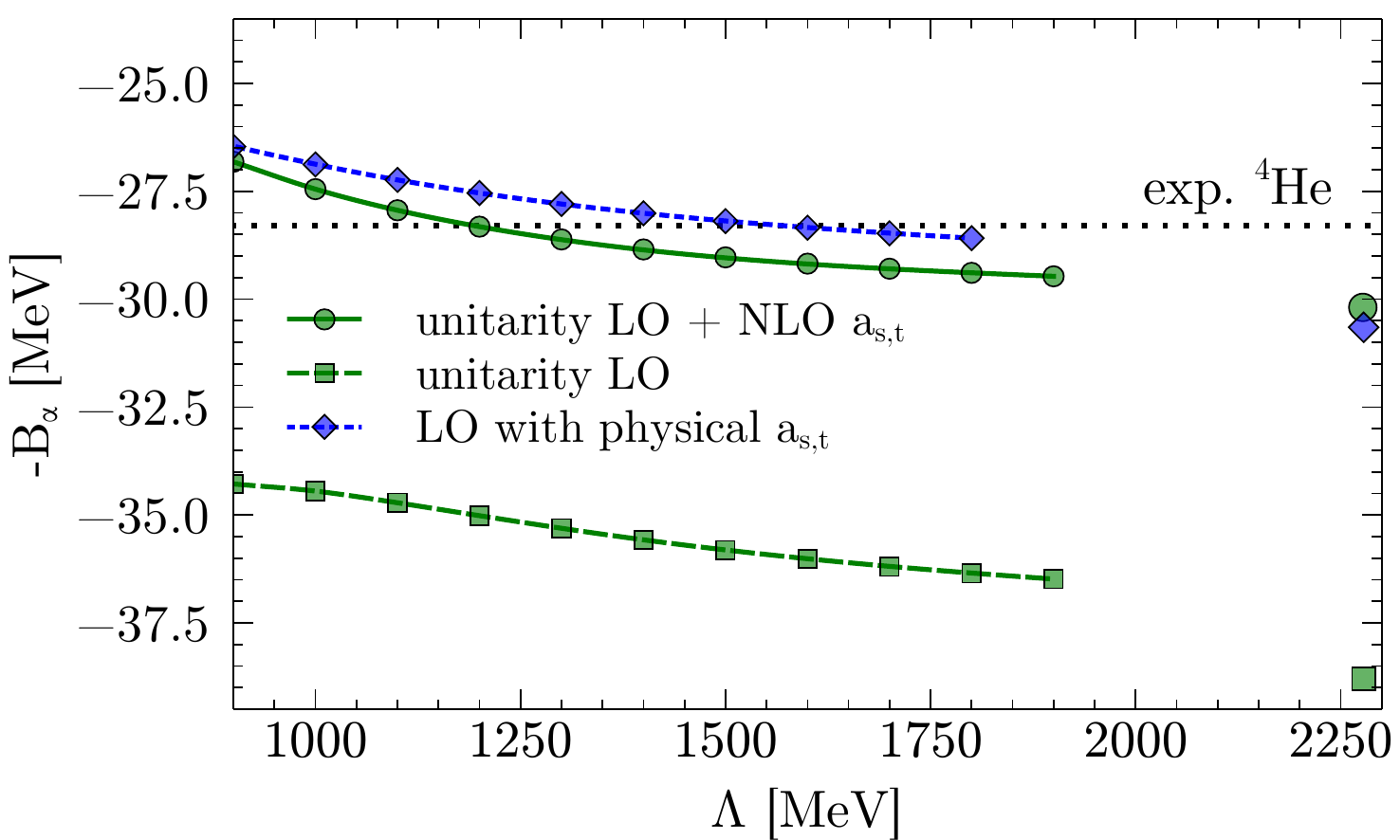}
\caption{%
\FourHe binding energy as function of the Gaussian cutoff parameter $\Lambda$.
The (blue) diamonds and (green) squares show, respectively, the results for
standard Pionless EFT and the unitarity expansion at leading order.
Inclusion of first-order corrections in $1/a_{s,t}$ (\ie, an incomplete \NLO
that neglects range and Coulomb effects) gives the (green) circles.
The closeness to the standard leading order demonstrates how well this part of
the expansion convergences.
Large symbols on the right edge indicate results for an extrapolation $\Lambda
\to \infty$ (see text).}
\label{fig:En-4He-gs}
\end{figure}

By construction, the unitarity expansion renders the deu\-te\-ron a zero-energy
bound state at leading order.
Since the expansion is set up in powers of the inverse scattering lengths,
it corresponds to the zero-range binding momentum $\kappa_t = 1/a_t$ in the
\ThreeSOne channel.
As demonstrated explicitly in Ref.~\cite{Konig:2016iny} by using the
perturbative formalism discussed in Sec.~\ref{sec:EFT}, this implies that
the deuteron remains at zero energy at \NLO and only moves to $1/(\MN
a_t^2)$ in an \NNLO calculation.
This is so for both the pure expansion in $1/a_{t}$, neglecting range
corrections, but interestingly also for the full unitarity expansion that
includes, \via $C_{2,\sti}^{(1)}$, range corrections starting at \NLO.
This is so because the unitarity LO shifts all corrections that mix ERE
parameters to a higher order compared to where they would be with a finite
scattering length at LO.
Overall, the dominant source of uncertainty for the deuteron energy comes from
the $1/(Q_2 a_t)$ expansion, which still amounts to a 50\% effect at \NNLO.
Conservatively taking the experimental binding energy as reference value for
the uncertainty estimate yields $B_d^\NNLO = 1.41 \pm 1.12~\MeV$.

\begin{table*}[t]
\centering
\begin{tabular}{c|c|c|c||c}
\hline\hline
\rule{0pt}{1.2em} state
& $E_B^\LO/\MeV$ & $E_B^\NLO/\MeV$ & $E_B^\NNLO/\MeV$ &
$E_B^{\text{exp.}}/\MeV$
\\
\hline
\rule{0pt}{1.2em}
\TwoH
\rule{0pt}{1.2em}
& $0$ & $0$ & $1.41 \pm 1.12$ & $2.22$\\
\ThreeH
& \underline{$8.48$} & \underline{$8.48$} & \underline{$8.48$} & $8.48$ \\
\ThreeHe
& $8.5 \pm 2.5$ & $7.6 \pm 0.2$ & \underline{$7.72$} & $7.72$ \\
\FourHe
& $39 \pm 12$ & $\;30 \pm 9^*$ & & $28.3$ \\
\hline\hline
\end{tabular}
\label{tab:EnergyResults}
\vspace{0.5em}
\caption{
Unitarity expansion convergence pattern.
Underlined values indicate energies which are used as input values to determine
three-body LECs.
An asterisk superscript indicates an incomplete \NLO result which only includes
the finite-scattering length but no contributions from effective ranges or
electromagnetic interactions.}
\end{table*}

\begin{figure}[tb]
\centering
\includegraphics[width=0.84\columnwidth,clip]{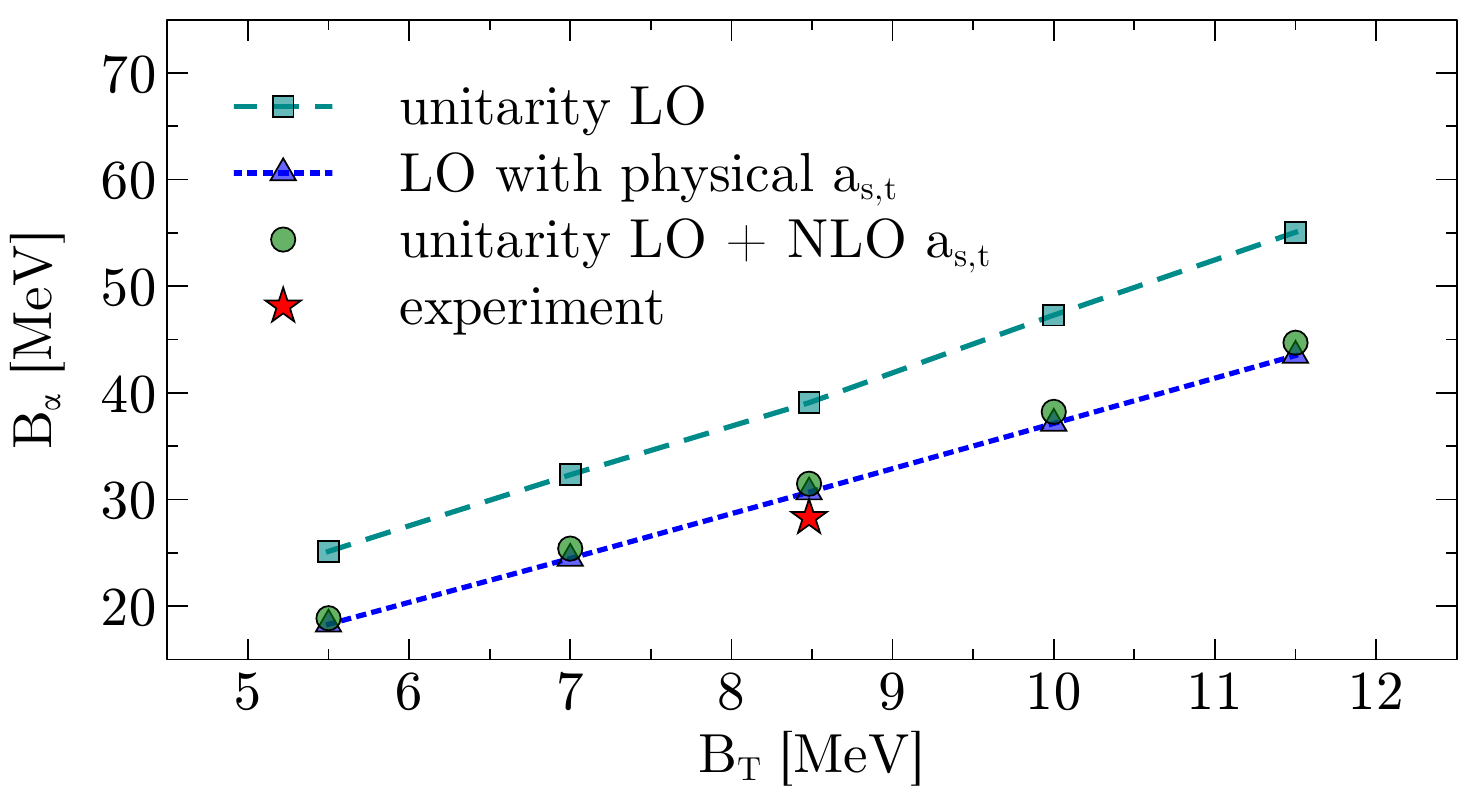}
\caption{%
Tjon line: correlation between the \FourHe and \ThreeH binding energies.
(Blue) dotted curve: standard pionless LO result; (green) dashed upper curve:
unitarity limit at LO.  Additional points nearly on top of the blue curve:
inverse scattering lengths added in first-order perturbation theory.
Star: experimental point.
}
\label{fig:TjonLine}
\end{figure}

The triton, being the ``anchor point'' of the expansion that determines the
value of three-body parameter $D_0^{(0)}$, stays fixed at the physical binding
energy at each order.
With the finite physical scattering lengths entering through
$C_{0,\sti}^{(1)}$ at NLO, the three-body LEC $D_0^{(1)}$ compensates the 
shift in the triton energy to keep it in place.
This leaves the \isotope[3]{He} binding energy as a nontrivial prediction.
While at LO by construction the trinucleon bound states are degenerate,
finite-scattering-length corrections together with Coulomb effects
(specifically, one-photon exchange) produce a triton-helion splitting
$(B_T-B_H)^\NLO \simeq (0.92 \pm 0.18)~\MeV$ at NLO~\cite{Konig:2016utl}.
Details regarding the perturbative treatment of Coulomb effects are discussed
in Refs.~\cite{Konig:2015aka,Konig:2016iny}.

The \FourHe nucleus provides a more serious test of the unitarity expansion.
Since $Q_A R \sim 0.8$ for \FourHe, it is the standard Pionless EFT part of the
unitarity scheme which naïvely one might doubt to work, while the pure unitarity
expansion, $\aleph/Q_A$, should indeed work better with increasing $Q_A$.
Figure~\ref{fig:En-4He-gs} shows the \FourHe binding energy as a function of
the momentum cutoff $\Lambda$.
The observed convergent behavior as $\Lambda$ increases indicates that the EFT
calculation is properly renormalized, as established originally in
Refs.~\cite{Platter:2005,Platter:2004he,Platter:2004zs}.
Results for the standard pionless LO, given by the (blue) diamonds in
Fig.~\ref{fig:En-4He-gs}, are consistent with this earlier work.

While any $\Lambda$ above the breakdown scale (of order $\Mpi$) is a valid
cutoff choice in principle, polynomials in $1/\Lambda$ are fitted to the points
in Fig.~\ref{fig:En-4He-gs} to quantitatively assess the convergence and
conveniently extrapolate $\Lambda\to \infty$.
This procedure gives $B_\alpha = 39(12)~\MeV$ in the unitarity limit, with the
uncertainty estimated as $\OO(r_{s,t}/a_{s,t}) \simeq 30\%$ based on the
expectation that range effects are the dominant correction.
Including the finite-scattering lengths as NLO corrections gives the (green)
circles in Fig.~\ref{fig:En-4He-gs}, very close to the standard pionless LO,
indicating that the $1/(Q_4a_{s,t})$ expansion appears to converge remarkably
well up to this order, and indeed the extrapolated result $30(9)$ $\MeV$ comes
out very near the standard pionless LO value of $31(9)~\MeV$.\footnote{
Note that the NLO data points at finite cutoffs shown in
Fig.~\ref{fig:En-4He-gs} differ slightly from the results shown in
Refs.~\cite{Konig:2016utl,Konig:2018ysz}.
This is due to a small error that has been fixed in the numerical
implementation.
Incidentally, the NLO result in the limit $\Lambda\to\infty$ is almost
unaffected by this, with the value merely changing from the $29.5~\MeV$ reported
in Ref.~\cite{Konig:2016utl} to $30.2~\MeV$.
The difference is negligible compared to the $\simeq9~\MeV$ uncertainty
estimated at this order.
Therefore, all conclusions of the previous work remain unchanged.}

It should be stressed that an NLO including only finite-$a$
corrections is incomplete: in the full unitarity expansion, range corrections
and Coulomb effects enter at the same time.
While the latter are expected to be small given that \FourHe is rather deeply
bound, it turns out that the inclusion of range effects actually has a profound
consequence:
in Ref.~\cite{Bazak:2018qnu} it is shown that a four-body interaction is
required to renormalize the universal four-boson system once range corrections
are included at \NLO, and universality implies that this conclusion carries over
directly to \FourHe in Pionless EFT.
The implication is that a four-nucleon input datum is required to fix the unknown
four-body parameter at \NLO, and this is most naturally taken to be
\FourHe ground-state energy.
Other properties, such as the ground-state radius (discussed below) or the
position of the $0^+$ excited state, will remain predictions at \NLO -- unless
it turns out that additional many-body forces are required for a renormalized
\NLO calculation of these observables.
While it may seem unlikely and is certainly not to be expected based on
NDA, such a possibility cannot \apriori be excluded:
each calculation needs to be carefully checked to be properly renormalized.

The rapid convergence of the pure unitarity expansion persists off the
physical point.
Figure~\ref{fig:TjonLine} shows the correlation between $3N$ and $4N$ binding
energies, known as the Tjon line~\cite{Tjon:1975sme}.
Its existence is explained by the three-body parameter largely governing the
physics of the system~\cite{Platter:2004zs}.
It is seen that the result starting from unitarity is shifted very close to
having the exact scattering lengths at LO over a significant range or triton
energies.
This observation provides further evidence that the unitarity expansion converges
well and that the results found at the physical point are not merely
accidental.

The unitarity expansion for ground-state energies up to $A=4$ is summarized in
Table~\ref{tab:EnergyResults}.  Observables fixed as input data at a given
order are shown as underlined text.  This in particular included the \ThreeHe
binding energy at \NNLO since according to
Refs.~\cite{Konig:2015aka,Konig:2016iny} an isospin-breaking three-body is
required once perturbation theory mixes Coulomb effects and range
corrections.\footnote{Note that if Coulomb effects are included
nonperturbatively already at leading order, which is not necessary for light
nuclei, an isospin-breaking three-body force enters already at
\NLO~\cite{Vanasse:2014kxa}.}  Taking into account the findings of
Ref.~\cite{Bazak:2018qnu}, the \FourHe binding energy should be underlined for
a complete \NLO as well.

\section{Charge radii and form factors}
\label{sec:Radii}

It has so far been established that the unitarity expansion describes well the
ground-state energies of light nuclei.
While certainly impressive given how simple the LO of the expansion is, it
is still merely a first step towards showing that the scheme comprehensively
captures the properties of light nuclei.

Further insight can be gained by considering charge radii of the $A=3,4$
systems as well.
Within the mo\-men\-tum-space framework employed in this work, this is achieved 
by
calculating charge form factors
\begin{equation}
 F_C(q^2) = \mbraket{\Psi}{\hat{\rho}(\vecq)}{\Psi}
\label{eq:FC}
\end{equation}
for the \ThreeH and \FourHe ground states, from which one obtains point charge
radii as
\begin{equation}
 \mean{r_0^2}
 = {-}\dfrac{1}{6}\dfrac{\dd}{\dd(q^2)} F_C(q^2)\Big|_{q^2=0}
 \mathtext{,}
 \mean{r_0} \equiv \sqrt{\mean{r_0^2}} \,.
\label{eq:r2-FC}
\end{equation}
In Eq.~\eqref{eq:FC}, the total charge operator $\hat{\rho} \equiv J_0$, \ie,
the zero component of the electric current $J_\mu$, is given by the sum of the
individual nucleon contributions,
\begin{equation}
 \hat{\rho} = \sum_{i=1}^A \hat{\rho}_i \,.
\end{equation}
The (anti-)symmetry of the wavefunction makes it possible to replace this sum by
$A\hat{\rho}_i$ for any fixed $i$.
A particularly convenient choice for three nucleons is $i=3$ because it holds
that
\begin{equation}
 \vecx_3 = \vecR^{(3)} + \frac23\vecr_2 \,,
\label{eq:x3}
\end{equation}
where $\vecr_2$ is the relative distance conjugate to $\vecu_2$ and
$\vecR^{(3)}$ is the overall center-of-mass coordinate.
With this choice, the momentum-space expression for the current operator
involves a momentum transfer only onto the Jacobi momentum $\vecu_2$.
Likewise, for four nucleons a good choice is $i=4$ because with
\begin{equation}
 \vecx_4 = \vecR^{(4)} + \frac34\vecr_3
\label{eq:x4}
\end{equation}
one obtains a momentum transfer only onto $\vecu_3$.

To use $\hat{\rho}$ within the Faddeev-Yakubovsky framework, it is necessary to
represent it within the appropriate partial-wave basis.
Since $\hat{\rho}$ does not depend on spin, two-body matrix elements of
$\hat{\rho}$ are given by
\begin{equation}
 \mbraket{u;\couple{l}{s}j m}{\hat{\rho}(\vecq)}{u';\couple{l'}{s'}j'm'}
 = \delta_{jj'}\delta_{mm'} \mbraketred{u;l}{\hat{\rho}(\vecq)}{u';l} \,,
\label{eq:rho-ujm}
\end{equation}
where the reduced matrix element on the right-hand side is given by
\begin{multline}
 \mbraket{u;l m}{\hat{\rho}(\vecq)}{u';l'm'}
 = \delta_{ll'}\delta_{mm'} \frac12\sum_{k}
 \sqrt{\binom{2l}{2k}} C_{k0,(l-k)0}^{l0} \\
  \null\times \int_{-1}^1 \dd x\,P_k(x)
  \frac{\delta\big(u'-\iota(u,q,x)\big)}{u'^2}
  \frac{u^{l-k}\big({-}\tfrac12q\big)^{k}}
  {\iota(u,q,x)^{l}} \,,
\label{eq:rho-ulm-final}
\end{multline}
without the $\delta_{mm'}$, where
\begin{equation}
 \iota(p,q,x) = \sqrt{\mathstrut p^2-pqx+q^2/4} \,.
\label{eq:iota}
\end{equation}
A detailed derivation of Eq.~\eqref{eq:rho-ulm-final} is provided in
Appendix~\ref{sec:ChargeOperators}.

Embedding $\hat{\rho}$ into the three- and four-body bases merely leads to
additional Dirac and Kronecker deltas, as well as to kinematic prefactors
multiplying the momentum transfer $q$ which can be read off from
Eqs.~\eqref{eq:x3} and~\eqref{eq:x4}:

\begin{strip}
\begin{equation}
 \mbraket{u_1u_2;s}{\hat{\rho}_3(\vecq)}{u_1'u_2';s'}
 = \delta_{l_1l_1'} \delta_{l_1l_1'} \delta_{s_1s_1'} \delta_{t_1t_1'}
  \delta_{s_2s_2'} \delta_{JJ'} \delta_{TT'}
  \frac{\delta(u_1-u_1')}{u_1^2}
  \mbraketred{u_2;l_2}{\hat{\rho}(\tfrac43\vecq)}{u_2';l_2'} \,,
\label{eq:rho-s}
\end{equation}
\begin{multline}
 \mbraket{
 u_1u_2u_3;a}{\hat{\rho}_4(\vecq)}{u_1'u_2'u_3';a'}
 = \delta_{l_1l_1'} \delta_{s_1s_1'} \delta_{j_1j_1'} \delta_{t_1t_1'}
  \delta_{s_2s_2'} \delta_{j_2j_2'} \delta_{t_2t_2'}
  \delta_{JJ'} \delta_{TT'}
  \frac{\delta(u_1-u_1')}{u_1^2} \frac{\delta(u_2-u_2')}{u_2^2}
  \mbraketred{u_3;l_3}{\hat{\rho}(\tfrac32\vecq)}{u_3';l_3'} \,.
\label{eq:rho-a}
\end{multline}
\end{strip}

\begin{figure}[htbp]
\centering
\includegraphics[width=0.99\columnwidth,clip]{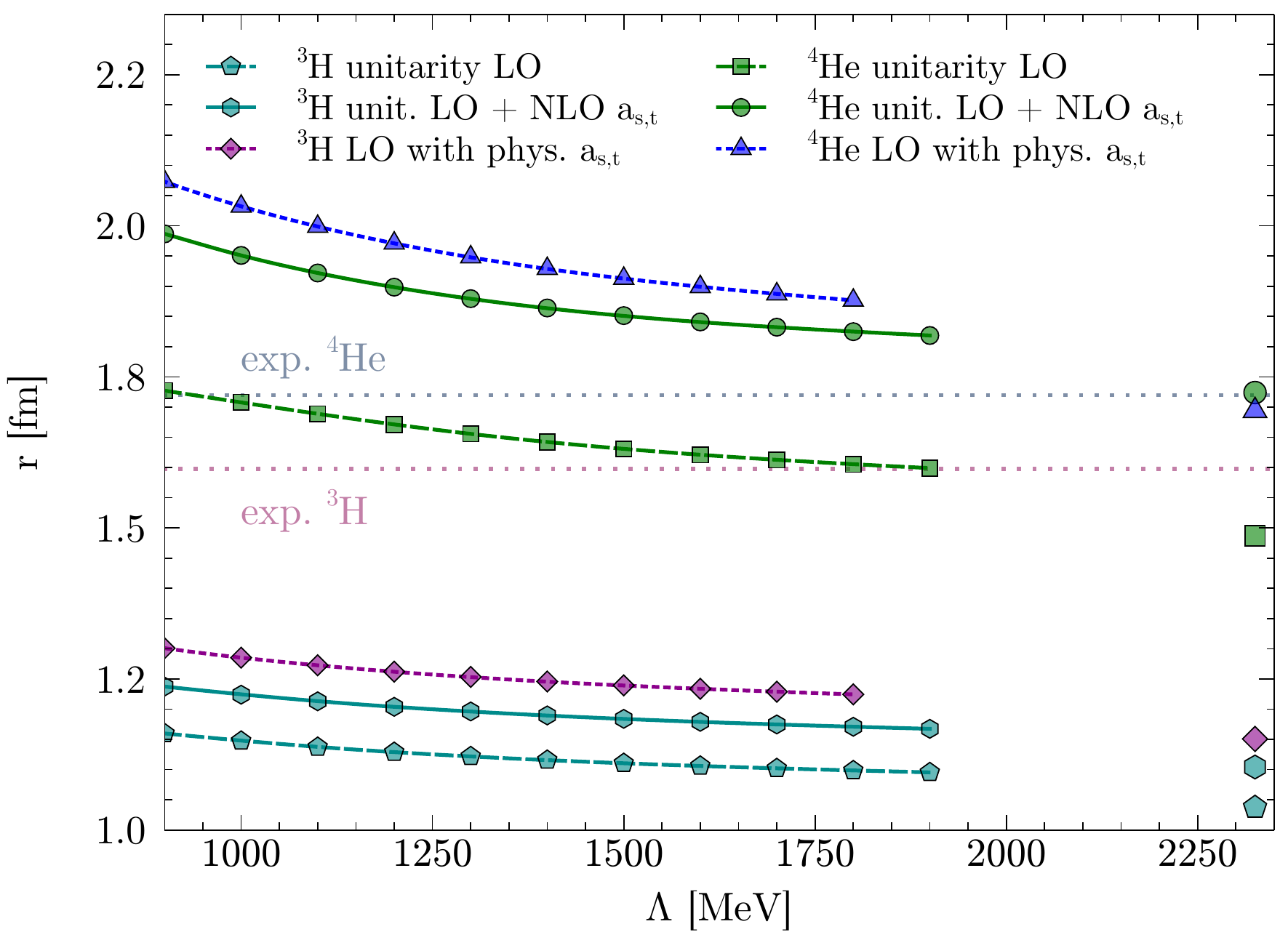}
\caption{%
Point-charge radii for \ThreeH and \FourHe as function of the Gaussian cutoff
parameter $\Lambda$.
The (purple) diamonds and (cyan) pentagons show, respectively, \ThreeH results
for standard Pionless EFT and the unitarity expansion at leading order.
The (cyan) hexagons are obtained by perturbatively including $1/\ast$
correction on top of the unitarity LO.
For \FourHe, results are shown in the upper part, with symbols as in
Fig.~\ref{fig:En-4He-gs}.
Large symbols on the right edge indicate results for an extrapolation $\Lambda
\to \infty$ (see text).}
\label{fig:Rad-3H-4He-gs}
\end{figure}

At leading order in the unitarity expansion, the form factor is given by
\begin{equation}
 F_C^{(0)}(q^2) = \mbraket{\Psi^{(0)}}{\hat{\rho}(\vecq)}{\Psi^{(0)}} \,,
\label{eq:FC-0}
\end{equation}
and analogously Eq.~\eqref{eq:r2-FC}, with added superscripts ``$(0)$,'' yields
the LO point charge radii.
At NLO, the correction to the form-factor is\footnote{
In principle, the current operator should also be expanded perturbatively,
$\hat{\rho} = \hat{\rho}^{(0)} + \hat{\rho}^{(1)} + \cdots$, but there is no
separate NLO contribution in the pure $1/a$ expansion considered here.
}
\begin{equation}
 F_C^{(1)}(q^2) = 2\,\mbraket{\Psi^{(1)}}{\hat{\rho}(\vecq)}{\Psi^{(0)}} \,,
\label{eq:FC-1}
\end{equation}
where the factor $2$ follows from symmetry.
Perturbatively expanding Eq.~\eqref{eq:r2-FC} gives
\begin{equation}
 \mean{r_0}^{(1)} = \frac12\frac{\mean{r_0^2}^{(1)}}{\mean{r_0}^{(0)}} \,,
\end{equation}
with $\mean{r_0^2}^{(1)}$ calculated from the slope of $F_C^{(1)}(q^2)$ at
$q^2=0$.

Evaluating the perturbative radius shifts defined above requires the NLO
correction to the wavefunctions that enter in Eq.~\eqref{eq:FC-1}.
Is is possible to obtain these for three- and four-body systems from
inhomogeneous versions of the Faddeev and Faddeev-Yakubovsky equations,
respectively.
As is derived in Appendix~\ref{sec:PerturbationTheory}, for three particles one
has $\ket{\Psi_1} = (1+P)\ket{\psi_1}$, with the NLO Faddeev
component $\ket{\psi_1}$ defined as a solution of\footnote{
Note that Eq.~\eqref{eq:Faddeev-psi1-3N} makes explicit use of the fact that the
three-body force considered here is symmetric under all permutations, such that
$V_3^{(n)}(1+P) = 3V_3^{(n)}$.}
\begin{multline}
 \left[1 - G_0 t^{(0)} P - (G_0 + G_0t^{(0)}\,G_0)V_3^{(0)}
 \right]\ket{\psi_1} \\
 = (G_0 + G_0 t^{(0)} G_0)\left[
 V_2^{(1)}(1+P) + V_3^{(1)}+ B_1
 \right]\ket{\psi_0} \,.
\label{eq:Faddeev-psi1-3N}
\end{multline}
In Eq.~\eqref{eq:Faddeev-psi1-3N}, $G_0$ and $t^{(0)}$ are understood to be
evaluated at the LO binding energy, $E = {-}B_0$.
Similarly, for four nucleons NLO Faddeev-Yakubovsky equations can be written as
\begin{equation}
 \left(\one - \hat{K}^{(0)}\right) \ket{\boldsymbol{\psi}^{(1)}}
 = \hat{K}^{(1)} \ket{\boldsymbol{\psi}^{(0)}} \,,
\label{eq:FY-1-mat}
\end{equation}
with the kernel $\hat{K}^{(0)}$ as defined in Eq.~\eqref{eq:K-FY-0-23} and
\begin{multline}
 \hat{K}^{(1)} = B_1 (G_0 + G_0t^{(0)}G_0) + G_0 t^{(1)} \hat{P} \\
 \null + G_0 t^{(1)} G_0 V_3^{(0)} \hat{P}_3
 + (G_0 + G_0t^{(0)}G_0) V_3^{(1)} \hat{P}_3 \,.
\label{eq:FY-K1-V3}
\end{multline}
Note that as explained in Appendix~\ref{sec:PerturbationTheory} special care
has to be taken when solving Eqs.~\eqref{eq:Faddeev-psi1-3N}
and~\eqref{eq:FY-1-mat} to account for the fact that the operators on the
left-hand sides are singular at $E={-}B_0$.
From the components $\ket{\boldsymbol{\psi}^{(1)}}
= (\ket{\psi_A^{(1)}}, \ket{\psi_B^{(1)}})^T$ the full NLO correction
$\ket{\Psi^{(1)}}$ is obtained analogously to Eq.~\eqref{eq:Psi-0-4B}.
For practical calculations Eq.~\eqref{eq:FY-1-mat} is simplified based on the
fact that all interactions are chosen to be separable.
This can be achieved with the same factorization as used in
Ref.~\cite{Platter:2005} at leading order.

As for the \FourHe energy discussed in the previous section, the focus here
is on the $1/a$ part of the unitarity expansion while the inclusion of
range corrections is postponed to future work.
Results for the ground-state radii of both \ThreeH and \FourHe are shown in
Fig.~\ref{fig:Rad-3H-4He-gs} as a function of the UV cutoff $\Lambda$.
Convergence as $\Lambda$ increases is evident from the plot, and just like it
was done for the binding energies polynomials in $1/\Lambda$ are fitted to
the data points in order to extrapolate $\Lambda\to\infty$.
The horizontal lines in Fig.~\ref{fig:Rad-3H-4He-gs} show the experimental
values of the point charge radii, which, following Ref.~\cite{Vanasse:2015fph},
are defined as
\begin{equation}
 \mean{r_0^2}_{\ThreeH}
 = \mean{r^2}_{\ThreeH} - \mean{r^2}_p - 2\mean{r^2}_n
\end{equation}
for the triton, and
\begin{equation}
 \mean{r_0^2}_{\FourHe}
 = \mean{r^2}_{\FourHe} - 2\mean{r^2}_p - 2\mean{r^2}_n
\end{equation}
for \FourHe.
That is, contributions from the root-mean-square radii of the individual
nucleons are subtracted from the experimental nuclear charge radii.

Using experimental values from Ref.~\cite{Angeli:2013xyz} for the quantities
appearing on the left-hand sides of the above definitions one obtains, with
error bars negligible compared to those of the present theoretical calculation,
$\mean{r_0^2}_{\ThreeH}^{\text{exp}} = 1.59~\fm$ and
$\mean{r_0^2}_{\FourHe}^{\text{exp}} = 1.72~\fm$.

The lower part of Fig.~\ref{fig:Rad-3H-4He-gs} shows results for the triton.
In the limit $\Lambda\to\infty$, indicated as points on the right border of the
plot, the \ThreeHe radius comes out as $\mean{r_0}_{\ThreeH}^{(0)} =
1.15(35)~\fm$ for the standard pionless LO, and $\mean{r_0}_{\ThreeH}^{(0)} =
1.04(31)~\fm$ at unitarity.
Perturbative corrections shift the unitarity LO result more than half way
towards the value obtained for physical scattering lengths at leading order,
\begin{equation}
 \mean{r_0}_{\ThreeH}^{(0)} + \mean{r_0}_{\ThreeH}^{(1)}
 = 1.10(33)~\fm \,,
\end{equation}
indicating that the unitarity expansion works well for this observable.
This is in line with the results of Ref.~\cite{Vanasse:2016umz}, where
good convergence is found for a perturbative expansion of \ThreeH and \ThreeHe
radii around an $SU(4)$ symmetric leading order (of which the unitarity limit is
a special case).
The result obtained for standard pionless LO is furthermore in excellent 
agreement with the calculation of Ref.~\cite{Vanasse:2015fph}, at unitarity
the radius satisfies well the universal
relation~\cite{Braaten:2004rn,Vanasse:2015fph,Vanasse:2016umz}
\begin{equation}
 \MN B_{\ThreeH} \mean{r_0^2}_{\ThreeH}
 = (1 + s_0^2)/9 \approx 0.224 \,.
\end{equation}

As done for binding energies it is assumed here that the $Q_AR$ part dominates
the overall expansion, yielding a $30\%$ uncertainty both at LO as well as NLO.
Indeed, from Ref.~\cite{Vanasse:2015fph} it is known that range corrections
contribute significantly to the triton radius in Pionless EFT and
shift the result close to the experimental value once they are included.
This uncertainty assignment places the experiment value outside the error band
of the unitarity LO result.
Since it is purely based on omitted range corrections, this is not
actually a reason for concern and merely indicates that to be yet more
conservative one should consider adding the uncertainties from the $Q_AR$ and
$\aleph/Q_A$ expansions coherently.

Results for \FourHe radius, shown in the upper part of
Fig.~\ref{fig:Rad-3H-4He-gs}, look equally good.
In fact, consistent with what is found for the binding energy, the result
obtained from a standard pionless LO calculation, $\mean{r_0}_{\FourHe}^{(0)} =
1.69(51)~\fm$, comes out surprisingly close to the experimental value.
At unitarity the radius comes out smaller, $\mean{r_0}_{\FourHe}^{(0)} =
1.49(45)~\fm$, consistent with the observed overbinding in the unitarity limit.
Inclusion of perturbative $1/a$ corrections shifts this value to
\begin{equation}
 \mean{r_0}_{\FourHe}^{(0)} + \mean{r_0}_{\FourHe}^{(1)}
 = 1.73(52)~\fm \,,
\end{equation}
in excellent agreement with the standard pionless LO result and therefore
providing yet more evidence for the good convergence properties of the
unitarity expansion.
In particular, the fact that convergence appears to be somewhat faster for
\FourHe than for the triton is in good agreement with $\aleph/Q_4 < \aleph/Q_3$
obtained from Eq.~\eqref{eq:Q-A}, and it therefore reinforces confidence in this
estimate.

\section{Summary and perspectives}
\label{sec:Outlook}

Superficially, the unitarity expansion may seem like merely a minor departure
from standard Pionless EFT.
It is rather well known that Pionless EFT, unlike Chiral EFT, is the
ideal EFT to describe few-nucleon systems at low energies since its expansion 
explicitly embraces implications from the scattering lengths
being large, basing its power counting explicitly on this fact.
Chiral EFT is limited at low energies by its simultaneous expansion in both 
momenta and around the chiral limit, with $\Mpi\neq0$ parametrizing the distance
from it.
This combination yields a power counting for $Q\sim\Mpi$ which does not easily 
capture the physics of the regime $Q\ll\Mpi$.
Notably, one-pion exchange only contributes to the $N\!N$ scattering lengths
through loop effects.

However, the unitarity expansion does in fact constitute a significant paradigm
shift in the EFT-based description of light nuclei: it goes as far as saying
that the details of the two-body sector, represented by the experimental
values of the scattering lengths, do not actually matter much to describe
properties of light nuclei.
Instead, it fully embraces universality and uses the three-body sector as
anchor point, constructing a leading order with just a single parameter and
an exact manifestation of the Efimov effect.
In this work it has been shown that the $1/a$ expansion of the unitarity 
scheme works well not only for binding energies of up to four nucleon systems,
but that similarly good convergence is obtain for the \ThreeH and \FourHe point
charge radii as well.
This finding solidifies the picture drawn in the introduction of this work,
placing few-nucleon systems in a universal regime perturbatively close to the
unitarity limit.

It is an important next step to include range corrections and Coulomb effects,
thus considering the full unitarity scheme that pairs the expansions in 
$\aleph/Q_A$ and $Q_A/\Mhi = Q_AR$.
So far, this has been investigated only for the \ThreeH-\ThreeHe energy
splitting, where by construction range corrections cancel at
\NLO~\cite{Konig:2016utl}.
An isospin-breaking three-nucleon force is required once range corrections
mix with Coulomb contributions at \NNLO~\cite{Konig:2016iny}.
Range corrections are known to significantly contribute to the triton point
charge radius~\cite{Vanasse:2015fph}.
For the \FourHe radius, the closeness to the experimental point already without
range corrections found in this work leaves little room for a significant shift 
at full \NLO.
From Ref.~\cite{Bazak:2018qnu} it is known that such a calculation will require
a four-nucleon force to be included, which is most conveniently fit to
reproduce the \FourHe energy at the experimental point at \NLO.
It is conceivable that this fixing of the energy will maintain a good
reproduction of the radius, just like it is observed for the standard Pionless
LO result.
Coulomb contributions should also be included in a complete \NLO
calculation, along with isospin breaking in the \OneSNot $NN$ scattering
lengths.
This is expected to be a small effect for a bound state as deep as \FourHe, but
it is interesting to note that fitting the four-nucleon force to exactly
reproduce the \FourHe binding energy at \NLO will inevitably absorb 
isospin-breaking contributions as well.
Clearly a careful overall consideration of the LEC fitting procedure is called
for in light of this to avoid possible overfitting of individual parameters.
Bayesian methods stand ready as a powerful tool to address
this~\cite{Wesolowski:2015fqa,Wesolowski:2018lzj}.

Apart from such more technical issues, it is an exciting question how far
into the nuclear chart the unitarity expansion can reach and what exactly its
place is in the tower of nuclear EFTs.
The observation that bosonic systems at unitarity exhibit saturation for large
numbers of particles~\cite{Carlson:2017txq} and recent calculations of nuclear
matter using interactions guided by unitarity~\cite{Kievsky:2018xsl} provide
reason to be optimistic that universality, and in particular discrete scale
invariance~\cite{vanKolck:2017zzf}, is able to inform more than just few-nucleon
calculations.
On the other hand, Refs.~\cite{Gattobigio:2019eqw,Dawkins:2019vcr} indicate
that few-nucleon systems beyond $A>4$ may not be bound in the unitarity limit.
To further assess this situation one should investigate whether these states
can be found as resonance (or virtual state) poles at unitarity, and if so, if
these poles are perturbatively close to the situation in the real world.

In the bosonic sector, the promotion of many-body forces to lower orders than
where they would be expected according to NDA is a fascinating consequence of
universality~\cite{Bazak:2018qnu}, but it does impose practical limitations on
many-body calculations.
Beyond four nucleons the influence of Fermi statistics is expected to become
important, which will most likely limit the promotion of $A$-nucleon 
forces with $A>4$.
However, at the same time one should wonder how much this might constrain the
usefulness of universality in general, and at which point the Fermi momentum 
becomes a relevant scale for the description of nuclei.
A calculation of, for example, $n$-$\alpha$ scattering within the unitarity
expansion will be an important next test of the framework and help
assess its exact place in the tower of nuclear EFTs.

\begingroup
\acknowledgement{%
I am grateful to Matthias Heinz for useful discussion about perturbation theory.
This work was supported by the Deutsche Forschungsgemeinschaft (DFG, German
Research Foundation) -- Projektnummer 279384907 -- SFB 1245 and by the
ERC Grant No.~307986 STRONGINT.  The numerical computations were performed
at the Jülich Supercomputing Center.
}
\endgroup

\appendix

\begin{strip}
\section{Partial-wave decomposition of charge operators}
\label{sec:ChargeOperators}

A generic one-body charge operator between two-body states $\ket{u;l m}$ can be
written as
\begin{spliteq}
 \mbraket{u;l m}{\hat{\rho}(\vecq)}{u';l'm'}
 &= \int\!\dd^3p\int\!\dd^3p'
  \,\braket{u;l m}{\vecp} \mbraket{\vecp}{\hat{\rho}(\vecq)}{\vecp'}
  \braket{\vecp'}{u';l'm'} \\
 &= \int\!\dd^3p\int\!\dd^3p'
  \,\YY_{l m}(\hat{p}) \frac{\delta(u-p)}{u^2}
  \vdelta\big(\vecp-\vecp'-\tfrac12\vecq\big)
  \YY_{l'm'}^*(\hat{p}') \frac{\delta(u'-p')}{u'^2} \\
 &= \int\!\dd p\,p^2\int\!\dd\Omega_{p}
  \,\YY_{l m}(\hat{p}) \frac{\delta(u-p)}{u^2}
  \frac{\delta(u'-\abs{\vecp-\tfrac12\vecq})}{u'^2}
  \YY_{l'm'}^*\big(\reallywidehat{\vecp-\tfrac12\vecq}\big) \,.
\label{eq:rho-ulm}
\end{spliteq}
The dependence on the angles of $\vecq$ can be isolated by using a procedure
analogous to the one described in Ref.~\cite{Gloeckle:1983} for the permutation
operators appearing in the Faddeev equations.
It holds that
\begin{equation}
 \YY_{l'm'}^*\big(\reallywidehat{\vecp-\tfrac12\vecq}\big)
 = \sum_{\lambda_1'+\lambda_2'=l'}
  \frac{p^{\lambda_1'}\big({-}\tfrac12q\big)^{\lambda_2'}}
  {\abs{\vecp-\tfrac12\vecq}^{l'}}
  \sqrt{\frac{4\pi(2l'+1)!}{(2\lambda_1'+1)!(2\lambda_2'+1)!}}
  \YYY_{\lambda_1'\lambda_2'}^{l'm'*}(\hat{p},\hat{q}) \,,
\end{equation}
and furthermore
\begin{equation}
 \frac{\delta(u'-\abs{\vecp-\tfrac12\vecq})}{u'^2}
 = 2\pi \sum_{k} \sqrt{\hat{k}} ({-}1)^k \left[
  \int_{-1}^1 \dd x\,P_k(x)
  \frac{\delta\big(u'-\sqrt{p^2-pqx+q^2/4}\big)}{u'^2}
  \right]
  \YYY_{kk}^{00}(\hat{p},\hat{q}) \,,
\end{equation}
where $\hat{k} = 2k + 1$.
\end{strip}

In these expressions, $\YYY_{l_1,l_2}^{LM}$ is used to denote two coupled
spherical harmonics.
The resulting product can be reduced:
\begin{multline}
 \YYY_{\lambda_1'\lambda_2'}^{l'm'*}(\hat{p},\hat{q})
 \YYY_{kk}^{00}(\hat{p},\hat{q})
 = \frac{1}{4\pi} \sqrt{\hat{k}\hat{\lambda_1'}\hat{\lambda_2'}}
  ({-}1)^{\lambda_1'+\lambda_2'+l'} \\
  \null\times
  \sum_{f_1,f_2}
  \SixJ{f_2}{f_1}{l'}{\lambda_1'}{\lambda_2'}{k}
  C_{k0,\lambda_1'0}^{f_10} C_{k0,\lambda_2'0}^{f_20}
  \YYY_{f_1f_2}^{l'm'*}(\hat{p},\hat{q})
 \,.
\end{multline}

At this point, it is possible to perform the integral over $\hat{p}$ in
Eq.~\eqref{eq:rho-ulm}, yielding:
\begin{equation}
 \int\!\dd\Omega_{p} \YY_{l m}(\hat{p})
 \YYY_{f_1f_2}^{l'm'*}(\hat{p},\hat{q}) \\
 = \sum_{m_2} C_{l m,f_2 m_2}^{l'm'} \YY_{f_2 m_2}^*(\hat{q})
  \,\delta_{f_1l} \,.
\end{equation}
Since $\hat{\rho}$ is a scalar operator that cannot connect $l' \neq l$,
only $f_2 = m_2 = 0$ can contribute and $\YY_{f_2 m_2}^*(\hat{q})$ reduced to a
factor $1/\sqrt{4\pi}$.
This leads to a cascade of further simplifications:
\begin{subalign}
 C_{l m,00}^{l'm'} &= \delta_{ll'}\delta_{mm'} \\
 \SixJ{0}{l}{l'}{\lambda_1'}{\lambda_2'}{k}
 &= ({-}1)^{l'+\lambda_1'+\lambda_2'}
  {\delta_{ll'}\delta_{\lambda_2'k}}
  / {\sqrt{\hat{l}\hat{k}}} \\
 C_{k0,k0}^{00} &= ({-}1)^k / \sqrt{\hat{k}} \,.
\end{subalign}
Putting everything together leads to Eq.~\eqref{eq:rho-ulm-final} in the main
text.

The next step is embedding the current into the three-nucleon system.
To that end one decouples the states $\ket{s}$ defined in Eq.~\eqref{eq:s} to
isolate the $\vecu_2$ part:
\begin{multline}
 \ket{u_1u_2;\couple{l_2}{\couple{\couple{l_1}{s_1}{j_1}}{\tfrac12}{s_2}}{JM}}
 \\
 = \sum_{m_2,\sigma_2}\sum_{\mu_1,\sigma}
  C_{l_2m_2,s_2\sigma_2}^{j_2\mu_2} C_{j_1\mu_1,\frac12\sigma}^{s_2\sigma_2}
 \\
 \null\times
 \ket{u_1;\couple{l_1}{s_1}{j_1}\mu_1}\ket{u_2;l_2m_2}\ket{\tfrac12\sigma} \,.
\end{multline}
Taking matrix elements, it is possible to exploit that $\hat{\rho}$ is
diagonal, so one simply gets a number of Kronecker deltas from the reduction of
the Clebsch-Gordan coefficients.
In particular,
\begin{equation}
 \braket{\couple{l_1}{s_1}{j_1}\mu_1}{\couple{l_1'}{s_1'}{j_1'}\mu_1'}
 = \delta_{j_1j_1'} \delta_{\mu_1\mu_1'} \delta_{l_1l_1'} \delta_{s_1s_1'} \,.
\end{equation}
For the isospin part, one finds
\begin{equation}
 \braket{\couple{t_1}{\tfrac12}{T}}{\couple{t_1'}{\tfrac12}{T'}}
 = \delta_{t_1t_1'} \delta_{TT'} \,,
\end{equation}
such that overall one arrives at Eq.~\eqref{eq:rho-s}.
Analogously, for the four-nucleon system, the first step is decoupling the
$\vecu_3$ part from the states~\eqref{eq:a}.
Omitting the intermediate quantum numbers for the $(123)$ subsystem as well as
the isospin part, one finds
\begin{multline}
 \ket{u_1u_2u_3;\couple{j_2}{\couple{l_3}{\tfrac12}{j_3}}{J}M} \\
 = \sum_{\mu_2,\mu_3}\sum_{m_3\sigma_3}
  C_{j_2\mu_2,j_3\mu_3}^{JM} C_{l_3m_3,\frac12\sigma_3}^{j_3\mu_3}
  \ket{u_1u_2;j_2\mu_2} \ket{u_3;l_3m_3} \ket{\tfrac12\sigma_3} \,,
\end{multline}
leading to Eq.~\eqref{eq:rho-a}.

\section{Perturbative expansion of few-body bound states}
\label{sec:PerturbationTheory}

This appendix discusses methods to obtain perturbative corrections for few-body
wavefunctions.
In the main text of this work, these corrections are used to calculate
first-order shifts for three- and four-nucleon charge radii within the unitarity
expansion, but having access to wavefunctions generally enables a variety of
further calculations.
For example, second-order corrections to binding energies
can be obtained in a way that is numerically much simpler than the procedure
used in Ref.~\cite{Konig:2016iny} to extract such second-order shifts from
off-shell $T$-matrix corrections.
An abstract operator notation is used throughout this section, and conventions
for sub- and superscripts differ from the usage in the main text in order to
simplify the notation.
After a general discussion that does not assume any fixed number of particles,
concrete first-order equations for three- and four-body states are derived in
Sec.~\ref{sec:PT-Faddeev}.
While explicit results are given for first-order perturbation theory, it is
clear from the following discussion that higher-order equations can be derived
analogously in a recursive fashion, much like it is done in
Ref.~\cite{Vanasse:2015fph} for a calculation of the triton charge radius up to
next-to-next-to leading order.
The derivations presented here are built on the concept of using inhomogeneous
equations to calculate perturbative corrections, introduced in
Ref~.\cite{Vanasse:2015fph} specifically for three-body vertex functions, and
previously in Ref.~\cite{Vanasse:2013sda} for scattering calculations.

\subsection{Generic discussion}
\label{sec:PT-generic}

In principle, one can base a perturbative corrections for bound-state
wavefunctions on the corresponding expansion of Lippmann-Schwinger equation
\begin{equation}
 T = V + V G_0 T \,,
\label{eq:LS-gen}
\end{equation}
following the discussion in Sec.~\ref{sec:EFT}.
Using the fact that as the energy approaches a bound-state at $E = {-}B$, the
$T$ matrix has a pole:
\begin{equation}
 T(E) = \frac{\ket{\Bgen}\bra{\Bgen}}{E + B}
 + \text{regular terms} \,.
\label{eq:T-pole}
\end{equation}
In this expression, the vertex function $\ket{\Bgen}$ is related to the
wavefunction $\ket{\Psi}$ via
\begin{equation}
 \ket{\psi} = G_0({-}B)\ket{\Bgen} \,.
\end{equation}
In order to develop a formalism that is connected to the bound-state Faddeev
formalism, it is however more instructive to start directly from the
Schrödinger equation.
Expanding
\begin{subalign}
 V &= V_0 + V_1 + \cdots \,, \\
 B &= B_0 + B_1 + \cdots \,, \\
 \ket{\Psi} &= \ket{\Psi_0} + \ket{\Psi_1} + \cdots \,.
\end{subalign}
gives the well-known first-order equation
\begin{equation}
 (H_0 + V_0) \ket{\Psi_1} + V_1 \ket{\Psi_0}
 = {-}B_0 \ket{\Psi_1} - B_1 \ket{\Psi_0} \,.
\label{eq:SG-Psi1-1}
\end{equation}
In principle, an explicit solution is given by
\begin{equation}
 \ket{\Psi_1} = \sum_{\alpha\in\mathcal{S}\setminus\{{-}B_0\}}
 \frac{\ket{\Psi_\alpha}\mbraket{\Psi_\alpha}{V_1}{\Psi_0}}
 {{-}B_0 - E_\alpha} \,,
\label{eq:Psi1-resolved}
\end{equation}
where $\mathcal{S}$ denotes the whole spectrum and the sum is generally a sum
over discrete states plus an integral over the continuous spectrum (note that
for simplicity it is assumed here that there are no degeneracies in the
spectrum).
Using Eq.~\eqref{eq:Psi1-resolved} however requires a complete diagonalization
of the leading-order Hamiltonian, which is not convenient for calculations
based on Faddeev- and Faddeev-Yakubovsky equations that by construction only
give access to specific individual states.
Moreover, for the systems considered in this work almost all contributions
in Eq.~\eqref{eq:Psi1-resolved} would come from discretized continuum of
scattering states, which is numerically challenging.
It is therefore interesting and relevant to look for alternative ways of
solving for $\ket{\Psi_1}$.

As a first step, one can rewrite Eq.~\eqref{eq:SG-Psi1-1} as
\begin{equation}
 \left[{-}B_0 - H_0 - V_0\right] \ket{\Psi_1}
 = \left[V_1 + B_1\right] \ket{\Psi_0} \,.
\label{eq:SG-Psi1-2}
\end{equation}
and recognize from this that Eq.~\eqref{eq:Psi1-resolved} involves the full
leading-order Green's function in spectral representation, with the bound-state
$\ket{\Psi_0}$ subtracted.
This subtraction is crucial because
\begin{equation}
 \left[{-}B_0 - H_0 - V_0\right] \ket{\Psi_0} = 0
\end{equation}
implies the operator on the right-hand side of Eq.~\eqref{eq:SG-Psi1-2} is
singular.
It is the term ${-}B_1\ket{\Psi_0}$ on the right-hand side that generates this
subtraction.
With this insight it is possible to derive methods to deal with the problem.\
Using the definition of the free Green's function,
\begin{equation}
 G_0(z) = (z - H_0)^{{-}1} \,,
\end{equation}
it is possible to further rewrite Eq.~\eqref{eq:SG-Psi1-2} as
\begin{equation}
 \left[1 - G_0(z)V_0\right] \ket{\Psi_1}
 = G_0(z) \left[V_1 + B_1\right] \ket{\Psi_0} \,.
\label{eq:SG-Psi1-final}
\end{equation}
with $z$ fixed at ${-}B_0$.
For any $z \neq {-}B_0$, the kernel on the left-hand side of
Eq.~\eqref{eq:SG-Psi1-final} is regular and can therefore be solved, after
discretization, as a linear system of equations.
Knowing from Eq.~\eqref{eq:Psi1-resolved} that the solution is actually well
defined in the limit $z\to{-}B_0$, numerical extrapolation can be used to
obtain $\ket{\Psi_1}$.

An alternative procedure is to consider a potential with the leading-order bound
state removed.
This can be achieved using the
replacement~\cite{Lehman:1982zz,Nogga:2005hy,Hlophe:2017bkd}
\begin{equation}
 V_0 \to V_0 + \lambda\ket{\Psi_0}\bra{\Psi_0}
\label{eq:V0-projected}
\end{equation}
in Eq.~\eqref{eq:SG-Psi1-final}, with $\lambda$ a large positive constant that
moves the bound state far away from the low-energy spectrum we are interested
in.
This procedure is valid because the orthogonality condition for the
first-order correction, $\braket{\Psi_1}{\Psi_0} = 0$, implies that the modified
equation is equivalent to original one.
Numerically, it gives excellent agreement with the extrapolation method for a
two-body test case.

\subsection{Faddeev and Faddeev-Yakubovsky decomposition}
\label{sec:PT-Faddeev}

\newcommand{\Vt}{\ensuremath{\tilde{V}}\xspace}
\newcommand{\Gt}{\ensuremath{G}\xspace}
\newcommand{\Tt}{\ensuremath{t}\xspace}

For a three-body state one can write, in analogy to Eq.~\eqref{eq:Psi-3},
$\ket{\Psi_1} = (1+P)\ket{\psi_1}$ and insert this decomposition into
Eq.~\eqref{eq:SG-Psi1-2}.
Three-body interactions are neglected here to keep the discussion as
transparent as possible.
Application of the remaining two-body interaction can be simplified by
exploiting the symmetry of the full wavefunction, which implies that
\begin{equation}
 V\ket{\Psi} = (1+P)\Vt\ket{\Psi} \,,
\end{equation}
where \Vt is the potential acting only on the specific pair of particles used
to define the Faddeev component $\ket{\psi}$.
This gives:
\begin{multline}
 \left[{-}B_0 - H_0 - (1+P)\Vt_0)\right] (1+P) \ket{\psi_1} \\
 = \left[(1+P)\Vt_1 + B_1\right] (1+P)\ket{\psi_0} \,.
\label{eq:SG-psi1-2a}
\end{multline}
Since $(1+P)$ commutes with the free Hamiltonian $H_0$, one can bring an
application of $(1+P)$ to the left of each term in this equation.
While $(1+P)$ can in general have zero eigenvalues, components that are in
$\ker(1+P)$ are clearly irrelevant for the description of the physical bound
state and it is therefore possible to proceed with a simplified equation:
\begin{multline}
 \left[G_0({-}B_0)^{{-}1} - \Vt_0\right]\ket{\psi_1} \\
 = \Vt_0P\ket{\psi_1}
 + \Vt_1(1+P)\ket{\psi_0} - B_1 \ket{\psi_0} \,,
\label{eq:SG-psi1-2b}
\end{multline}
where $G_0({-}B_0)^{{-}1} = {-}B_0 - H_0$.
Acting with $G_0({-}B_0)$ on both sides of this one can perform a partial
inversion by using the Lippmann-Schwinger equation
\begin{equation}
 \left[1 - G_0 \Vt_0\right]\Gt = G_0
\end{equation}
for the full two-body Green's function \Gt associated with \Vt.
This can then be eliminated in favor of the corresponding $t$ matrix by using
\begin{subalign}
 \Gt\Vt &= G_0\Tt \,, \\
 \Gt &= G_0 + G_0 \Tt G_0 \,,
\end{subalign}
for $t = \Vt + \Vt G_0 t$.
Overall one arrives at:
\begin{equation}
 \left[1 - G_0 \Tt_0 P\right]\ket{\psi_1}
 = (G_0 + G_0 \Tt_0 G_0)\left[\Vt_1(1+P) + B_1\right] \ket{\psi_0}
\label{eq:SG-psi1-final}
\end{equation}
with $G_0 = G_0({-}B_0)$.
This is an inhomogeneous integral equation which involves exactly the same
kernel as the Faddeev equation at leading order, whereas the terms on the
left-hand side are straightforward to calculate from known quantities.
As a final step one calculates $\ket{\Psi_1}$ from $\ket{\psi_1}$ in exactly
the same way $\ket{\Psi_0}$ is calculated from $\ket{\psi_0}$.
By keeping three-body forces in the derivation one analogously obtains
Eq.~\eqref{eq:Faddeev-psi1-3N} given in the main text.

It is useful to note that the same result can alternatively be derived directly
from a perturbative expansion of the Faddeev equation
\begin{equation}
 \ket{\psi} = G_0 \Tt P \ket{\psi} \,,
\end{equation}
setting $\ket{\psi} = \ket{\psi_0} + \ket{\psi_1} + \cdots$.
To proceed it is important to carefully regard the energy arguments of the
operators, noting that $B = B_0 + B_1 + \cdots$:
\begin{subalign}
 \label{eq:G0-expansion}
 G_0({-}B) &= G_0({-}B_0) + B_1 G_0({-}B_0)^2 + \cdots \,, \\
 \label{eq:t-expansion}
 t({-}B) &= t_0({-}B_0) - B_1 \frac{\dd}{\dd z} t_0(z) \Big|_{z={-}B_0}
 + t_1({-}B_0) + \cdots \,.
\end{subalign}
The term involving the energy derivative in Eq.~\eqref{eq:t-expansion} looks
peculiar, but, assuming the potential to be energy independent, it can be shown
by differentiating the Lippmann-Schwinger equation that
\begin{equation}
 \frac{\dd}{\dd z} t_0(z) = {-}t_0(z) G_0(z)^2 t_0(z) \,.
\end{equation}
Using this and making use of the leading-order Faddeev equation,
\begin{equation}
 \ket{\psi_0} = G_0({-}B_0) \Tt_0({-}B_0) P \ket{\psi_0} \,,
\label{eq:Faddeev-psi0}
\end{equation}
to simplify some terms, one obtains
\begin{equation}
 \left[1 - G_0 \Tt_0 P\right]\ket{\psi_1}
 = B_1 (G_0 + G_0 \Tt_0 G_0) \ket{\psi_0} + G_0 \Tt_1 P \ket{\psi_0}
\label{eq:SG-psi1-alt}
\end{equation}
All energy arguments are ${-}B_0$ and have therefore been omitted as well.
Using again Eq.~\eqref{eq:Faddeev-psi0} as well as the Lippmann-Schwinger
equation for $\Tt_1$, Eq.~\eqref{eq:LS-t1} in the main text, it possible to
show that
\begin{equation}
 G_0 \Tt_1 P \ket{\psi_0} = (G_0 + G_0 \Tt G_0)\Vt_1(1+P) \ket{\psi_0} \,,
\end{equation}
so that indeed Eq.~\eqref{eq:SG-psi1-alt} is equivalent to
Eq.~\eqref{eq:SG-psi1-final}

\medskip
The previous result is both reassuring and very useful because it implies that
in order to derive perturbative corrections for four-body states one can simply
start from the Faddeev-Yakubovsky equations, avoiding a tedious detour through
the Schrödinger equation.

A convenient starting point for this calculation is generic matrix form of the
Faddeev-Yakubovsky equations, analogous to Eq.~\eqref{eq:FY-0-mat}:
\begin{equation}
 \left(\one - \hat{K}\right) \ket{\boldsymbol{\psi}} = 0 \,,
\end{equation}
with $\ket{\boldsymbol{\psi}} = (\ket{\psi_A}, \ket{\psi_B})^T$ and
\begin{equation}
 \hat{K} = G_0 \Tt \hat{P} \,,
\label{eq:K-FY-gen}
\end{equation}
where three-body forces are again neglected for simplicity.

As before, one expands all quantities in this equation, carefully keeping track
of the energy arguments.
This gives
\begin{equation}
 \left(\one - \hat{K}_0\right) \ket{\boldsymbol{\psi}_1}
 = \hat{K}_1 \ket{\boldsymbol{\psi}_0}\,,
\label{eq:FY-psi-final}
\end{equation}
with
\begin{subalign}[eq:K-FY-01]
\label{eq:K-FY-0}
 \hat{K}_0 &= G_0 \Tt_0 \hat{P} \,, \\
\label{eq:K-FY-1}
 \hat{K}_1 &= B_1 (G_0 + G_0 \Tt_0 G_0) + G_0 \Tt_1 \hat{P} \,.
\end{subalign}
All energy arguments in Eqs.~\eqref{eq:K-FY-01} are fixed now at ${-}B_0$ and
have been omitted.
Again the term involving $\Tt_1$ can be rewritten in terms of $\Vt_1$ as
\begin{equation}
 G_0 \Tt_1 \hat{P} \ket{\boldsymbol{\psi}_0}
 = (G_0 + G_0 \Tt_0 G_0)\Vt_1(1+\hat{P}) \ket{\boldsymbol{\psi}_0} \,.
\end{equation}
With three-body forces included one obtains the slightly more complicated
$\hat{K}_1$ shown in Eq.~\eqref{eq:FY-K1-V3} in the main text.

\medskip
The fact that within the Faddeev(-Yakubovsky) formalism one has to calculate
$B_1$ with an increasing number of partial-wave components included to check
convergence of the numerical calculation renders the extrapolation method
discussed in Sec.~\ref{sec:PT-generic} unstable beyond the two-body sector.
At the same time, the projection method which modifies the potential to remove
the leading-order bound state from the low-energy spectrum becomes impractical
because for an $A$-body state Eq.~\eqref{eq:V0-projected} involves an $A$-body
potential, which is expensive to handle computationally (at leading order) for
$A>3$.
Fortunately, it is possible to employ the projection method directly at the
level of Faddeev(-Yakubovsky) components and make the replacement\footnote{
Alternatively, one can work directly with the singular $K_0$ and solve
the first-order equation as a linear least-squares problems using a singular
value decomposition (SVD).}
\begin{equation}
 K_0 \to K_0 + \lambda \ket{\psi_0}\bra{\psi_0}
\end{equation}
in Eq.~\eqref{eq:SG-psi1-final}, where $K_0 = G_0 t_0 P$, and analogously
for Eq.~\eqref{eq:FY-psi-final}.
While this ensures that the kernel becomes regular, the solution of the
modified equation, denoted by $\ket{\tilde{\Psi}_1}$, will not in general
directly provide the correct solution $\ket{\Psi_1}$ (obtained from
$\ket{\psi_1}$ by adding the appropriate permutations).
In particular, $\ket{\tilde{\Psi}_1}$ may not be orthogonal to $\ket{\Psi_0}$.
It is however possible to simply project out the undesired component by setting
\begin{equation}
 \ket{\Psi_1}
 = \ket{\tilde{\Psi}_1} - \ket{\Psi_0}\braket{\tilde{\Psi}_1}{\Psi_0} \,,
\end{equation}
which gives the desired the solution.
The correctness of the result can be checked by evaluating
\begin{equation}
 \mbraket{\Psi_1}{H_0 + V_0}{\Psi_1} + \mbraket{\Psi_1}{V_1}{\Psi_0}
 = {-}B_0 \braket{\Psi_1}{\Psi_1} \,,
\end{equation}
which follows from Eq.~\eqref{eq:SG-Psi1-1}, since $B_0$ is known from the
leading-order calculation.

\end{document}